\documentclass{article}
\usepackage[margin=2cm]{geometry}

\PassOptionsToPackage{numbers, compress}{natbib}

\PassOptionsToPackage{prologue,dvipsnames}{xcolor}

\usepackage[utf8]{inputenc} 
\usepackage[T1]{fontenc}    
\usepackage[hidelinks]{hyperref}       
\usepackage{url}            
\usepackage{booktabs}       
\usepackage{amsfonts}       
\usepackage{nicefrac}       
\usepackage{microtype}      

\let\oldemph\emph
\renewcommand{\emph}[1]{{\color{BrickRed}\oldemph{#1}}}

\usepackage{wrapfig}

\usepackage{amsmath}
\usepackage{amssymb}
\usepackage{mathtools}
\usepackage{amsthm}
\usepackage{bbm}

\usepackage[capitalize,noabbrev,nameinlink]{cleveref}

\theoremstyle{definition}
\newtheorem{theorem}{Theorem}

\theoremstyle{remark}

\usepackage[textsize=tiny]{todonotes}

\usepackage{booktabs}       
\usepackage{nicefrac}       
\usepackage{microtype}      
\usepackage{xcolor}         
\usepackage{xspace}

\usepackage{comment}
\usepackage[T1]{fontenc}

\usepackage{tikz}
\usetikzlibrary{patterns}
\usetikzlibrary{calc}
\usetikzlibrary{arrows.meta}
\usetikzlibrary{positioning}
\usetikzlibrary{shapes.geometric}
\usetikzlibrary{decorations.pathreplacing}
\usepackage{pgfplots}
\pgfplotsset{compat=1.18}

\DeclareMathOperator{\prm}{\mathsf{PR}}
\DeclareMathOperator{\rwd}{\mathsf{RWD}}
\DeclareMathOperator{\evw}{\mathsf{EVW}}
\DeclareMathOperator{\f}{\mathsf{F}}

\usepackage{mathtools}

\newcommand{\absorbdecay}{{\textsc{MaxMinAbsorb}}\xspace}

\newcommand{\topdecay}{{\textsc{TopDecay}}\xspace}
\newcommand{\toprank}{{\textsc{TopRank}}\xspace}

\usepackage[T1]{fontenc}

\newcommand{\Succ}{S\!ucc}
\newcommand{\Pred}{Pred}

\usepackage{pifont}
\definecolor{crimson}{RGB}{220, 20, 60}
\definecolor{darkgreen}{rgb}{0,0.5,0}

\definecolor{mossgreen}{rgb}{0.68, 0.87, 0.68}

\usepackage{amsthm}

\usepackage{enumitem}
\setlist[description]{topsep=0pt, partopsep=0pt, itemsep=2pt, leftmargin=2em}

\usepackage{cleveref}

\newcommand{\E}{\ensuremath{\mathbbm{E}}} 
\newcommand{\CalD}{\ensuremath{\mathcal{D}}} 
\renewcommand{\P}{\ensuremath{\mathbbm{P}}} 
\newcommand{\1}{\ensuremath{\mathbbm{1}}}
\newcommand{\D}{\ensuremath{G}} 
\newcommand{\N}{\ensuremath{V}} 

  \tikzset{     
    e4c node/.style={circle,draw,minimum size=0.3cm,inner sep=0,font=\tiny},
    e4b node/.style={circle,draw,minimum size=0.8cm,inner sep=0,font=\normalsize},
    selected/.style={draw=BrickRed,double}, 
    selected2/.style={preaction={fill, OliveGreen!45}}, 
    selected3/.style={}, 
    e4c edge/.style={sloped,above,font=\footnotesize,-{Classical TikZ Rightarrow[length=0.75mm]},draw,thin},
    e4c path/.style={-{Classical TikZ Rightarrow[length=0.75mm]},draw,thin}
  }

\tikzset{
  efc-node/.style={
    circle,
    draw=black,
    line width=0.6pt,
    minimum size=0.70cm,
    inner sep=0pt,
    fill=white,
    font=\fontsize{5}{5.5}\selectfont
  },
  efc-edge/.style={
    ->,
    draw=black,
    line width=0.5pt
  }
}
 
\newcommand{\efcSL}[2]{%
  \renewcommand{\arraystretch}{1.1}%
  \begin{tabular}{@{}c@{}}
{\textcolor{BrickRed}{#1}} \\[1.5pt]
    \hline\\[-3.5pt]
    #2
  \end{tabular}%
}
 
\tikzset{>=stealth}
\title{
Power in Liquid Democracy: A Network Centrality Approach}

\usepackage{natbib}
\usepackage{authblk}

\author{
Davide Grossi$^{1,2}$ \quad
Andreas Nitsche$^{3}$ \quad
Georgios Papasotiropoulos$^4$ \\
Oskar Skibski$^{4}$ \quad
Piotr Skowron$^{4}$ \quad
Tomasz Wąs$^5$\\
\vspace{0.6cm}
{\small
$^1$University of Groningen\par
$^2$University of Amsterdam\par
$^3$Association for Interactive Democracy\par
\vspace{1pt}
$^4$University of Warsaw\par
\vspace{-2pt}
$^5$University of Oxford
}
}
  \date{}

\begin{document}

\maketitle

\begin{abstract}
This paper develops a computationally tractable framework for measuring voters’ power in collective decision-making platforms that support transitive and suspendable delegations—commonly known as liquid democracy. We employ Random Walk Decay centrality to capture how influence propagates through delegation networks, argue for its intuitive appeal in this context and its advantages over alternatives such as PageRank, and derive a natural axiomatic characterization within the class of power metrics. Moreover, we conceptualize how the framework can be extended into a practical tool for analyzing power distributions in such platforms, and propose and study, both axiomatically and algorithmically, methods for selecting representative slates of influential participants.\looseness-1
\end{abstract}

\section{Introduction}
\label{sec:intro}
\emph{Liquid democracy} (LD) is a framework for flexible participation in collective decision-making that can be used across multiple issues over time. One of its characteristic features \cite{blum2016liquid,valsangiacomo2022clarifying} is the suspendible delegation of voting rights: participants may delegate their voting rights to a representative, while retaining the ability to cast a direct vote at any time, thereby suspending the delegation. Delegations in LD can therefore be viewed as conditional instructions: representatives act on behalf of participants only on issues for which the latter do not cast direct votes. Representatives may issue the same kind of instruction and delegate onwards, and, if they do not vote directly, this onward delegation also passes on the delegations they have received. This leads to the defining feature of delegations in LD: their transitivity. Whoever acts as the  representative of my representative, also acts as my own representative. A voter then casts a vote for themselves and all the participants that delegated to them, directly or indirectly.\looseness-1

The framework is deployed in real-world decision-making scenarios through software implementations \cite{paulin2020overview} such as, in particular, \emph{LiquidFeedback}\footnote{\url{https://liquidfeedback.com}}---the decision-making platform that directly inspires this work. In LiquidFeedback the conditional nature of delegation is operationalized by automatically suspending outgoing delegations whenever a participant acts directly 
\cite{behrens2014principles}.
Delegative voting has also been extensively studied and implemented in the context of governance within decentralized autonomous organizations (DAOs); refer to \cite{weidener2025delegated} for a systematic review.

The transitivity of delegations guarantees a form of flexible representation of participants, while maintaining the one-person-one-vote principle that is typical in direct democracy. At the same time, it has given rise to concerns about the possibility of disproportionate accrual of power, thereby harming the democratic legitimacy of LD-enabled decisions. The aim of this paper is to provide a computationally tractable and quantitative theory of the way in which the transitivity of delegations impacts power, or influence, in decision making via LD. The significance of such a theory is twofold. 

Conceptually, it provides solid foundations from which to arbitrate claims concerning possible negative effects of the feature of transitivity on the concentration of power in LD systems. Practically, it enables much-needed on-the-fly diagnostics of power distributions during LD-based collective decisions.\looseness-1

To date, power in LD is an understudied topic, as also emphasized in the recent survey of the field \cite{ldsurvey}. In \cite{kling15voting} the authors provided an empirical analysis of power and influence based on data from LiquidFeedback.
Later on, a first formal theory of power was proposed in \cite{zhang2021power}, and subsequently extended in \cite{colley23measuring}. While such a theory managed to capture some intuitive features of power in LD, it had the important drawbacks of being computationally intractable and more cumbersome to analyze, due to its reliance on adaptations of the Penrose-Banzhaf index \cite{banzhaf,penrose46elementary}. Instead of using the machinery of power indices and voting games, in this paper we leverage and adapt tools from Network Science \cite{newman2018networks}, in particular, centrality measures \cite{shvydun2025zoo}. 
Connections between LD and centralities have already been explored in the literature. Specifically, PageRank, Katz and eigenvector centralities have been considered as a voting-weight determination method for casting voters, in delegation-based systems for elections \cite{armstrong2024optimizing,bersetche2025generalizing,boldi2009voting,boldi2011viscous} and DAOs \cite{hartnell2025verifiable,leifman2014secret}. In contrast, we study influence at the level of all participants—not only casting voters—and advocate for the Random Walk Decay centrality \cite{was2019random}.

From the modeling perspective, we closely follow the implementation used in LiquidFeedback \cite{behrens2014principles}, a model that was also very recently studied from a game-theoretic perspective by \citet{brillcycles}. 
In contrast, we provide an axiomatic perspective in the spirit of the works by \citet{utke2023anonymous,brill2022liquid} and \citet{boldi2014axioms}.
We also study, as an application of our power determination metric, the problem of identifying important users in a delegation network, that has been referred to in the literature as the peer selection problem \cite{aziz2016strategyproof,lev2024impartial,olckers2024manipulation,papasotiropoulos2025proportional}.

\smallskip
\textbf{Our Contributions.}
The paper makes the following contributions. {\em First}, it defines a novel metric of power in LD systems that incorporates the suspendibility of delegations: \emph{expected voting weight} (EVW) of users. The metric is conceptually simple and computationally tractable, in contrast to the existing approaches in the relevant literature, while retaining intuitive appeal for applications such as LiquidFeedback. {\em Second}, it shows that the proposed power metric coincides on delegation graphs with the \emph{Random Walk Decay} (RWD) centrality measure. This establishes delegative democracy as a particularly well-suited application domain for RWD, for which no canonical use-case has yet been identified.
Moreover, unlike more commonly studied metrics such as PageRank—which we show to be considerably less fitting in the present context—RWD has received virtually no attention in the computational social choice literature~\cite{Handbook-COMSOC}. {\em Third}, it articulates a strong argument in favor of EVW/RWD, by characterizing it as the unique power metric that satisfies a set of insightful, domain-specific axioms tailored to delegation systems. {\em Finally}, by leveraging the above theory, we apply it to the problem of \emph{proportional peer selection}.
The paper advances a concrete proposal for an algorithm that can identify a slate---that is, a set of participants of a given size---which effectively represents the structure that emerged from a voting platform during a delegation-based decision process. We argue that, once implemented in platforms like LiquidFeedback, the algorithm could further improve the transparency, effectiveness, and versatility of the collective decision-making process for users.
We evaluate the proposed method both axiomatically and algorithmically.\looseness-1

\section{Model and Terminology}

\label{sec:prelims}

The paper operates within the \emph{model of suspendible delegations}, which closely reflects the implementation used in LiquidFeedback \cite{behrens2014principles} and was previously studied by \citet{brillcycles}.
Throughout the paper, we interpret nodes of a given graph as voters (or agents/users) in a delegative democracy setting, and edges as expressions of delegation. We refer to such graphs as \emph{delegation graphs}.
In particular, we view this graph as a “potential-delegation graph” which is reused across multiple elections, towards making decisions on different issues. A voter initially specifies one or several outgoing edges, indicating their \emph{default delegation}, i.e.,
where their vote should be delegated in case they choose not to vote directly for some issue. For certain issues or topics, the voter may still cast a direct vote, in which case no representative is needed. Therefore, the delegation option acts as a fallback---one can say that delegations can be \emph{suspended} at any time if the agent decides to vote directly. 
In what follows, unless stated otherwise, we focus on scenarios in which each voter delegates to at most one other voter, the so-called \emph{single-proxy} setting. This setting, not only constitutes the main case considered in practice \cite{behrens2014principles} but it is also the most commonly studied model in the literature of LD---for a complete list of relevant works, we refer to the entries marked as “single-proxy” on the website accompanying the survey by \citet{ldsurvey}.

A \emph{directed graph}, or a \emph{network} is defined as a pair $G = (V,E)$, where $V$ is a set of $n$ nodes, or nodes, and $E \subseteq V \times V$ is a set of $m$ (directed) edges, i.e., ordered pairs of nodes.
For an edge $(u,v)$, we say that it is \emph{outgoing} from $u$ and \emph{incoming} to $v$.
We denote by $E^+(u)$ the set of edges outgoing from~$u$, and for a subset $S \subseteq V$, we write $E^+(S) = \bigcup_{u \in S} E^+(u)$.
The number of outgoing (respectively, incoming) edges of a node $u$ is called its \emph{out-degree} (respectively, \emph{in-degree}) and is denoted by $\deg^+(u)$ (respectively, $\deg^-(u)$). 
Any node $u$ such that $\deg^+(u)=0$ will be called \emph{endpoint}.
A graph $G = (V,E)$ is called \emph{functional} if every node has out-degree at most one. This terminology reflects that the edge set can be viewed as defining a (partial) function assigning to each node at most one other. When convenient, we will assume that $\deg^+(u)=1$ for all $u \in V$ by introducing self-loops.\looseness-1

A \emph{walk} in $G$ is a sequence of nodes $(v_1,\dots,v_k)$ such that $(v_i,v_{i+1}) \in E$ for every $i \in \{1,\dots,k-1\}$.
Its \emph{length} is the number of edges it contains, namely $k-1$. In particular, a single node forms a walk of length $0$.
If all nodes in the sequence are distinct, the walk is called a \emph{path}.
We use $\Omega(G)$ to denote the collection of all walks in $G$.
We say that a node $u$ is a \emph{predecessor} of a node $w$ if there exists a walk of positive length starting at $u$ and ending at $w$. Conversely, $u$ is a \emph{successor} of $w$ if $w$ is a predecessor of $u$.
The sets of predecessors and successors of a node $u$ are denoted by $\Pred_G(u)$ and $\Succ_G(u)$, respectively, omitting subscripts when clear from the context.
We will also refer to the \emph{(voting) weight} $w_u$ of $u$ as the value $1+|\Pred(u)|$.
For a subset of edges $M \subseteq E$, we write $G - M$ for the graph obtained by removing these edges, i.e., $(V, E \setminus M)$.
A \emph{(weakly) connected component} of $G$ is a maximal subset of nodes $S \subseteq V$ such that every pair of nodes in $S$ is connected by a walk in the underlying undirected graph of $G$, which is the graph that emerges after disregarding the directions of edges.
A \emph{clique} is a graph in which for every pair of nodes $u,v$ the graph contains the edge $(u,v)$.

We will use $p_G(u) \in \mathbb{R}^+$ to refer to the \emph{power} of a node $u$ in the graph $G$, which we will mainly measure in terms of a centrality index.
A \emph{centrality measure} $F$ is a function that for each $G=(V,E)$ and $v \in V$ assigns a real value, denoted by $F_G(v)$, to reflect the importance of $v$ in $G$.
\emph{PageRank}~\citep{PagBriMotWin-1999-PageRank} ($\prm$) and \emph{Random Walk Decay}~\cite{was2019random} ($\rwd$) are centrality measures which, given a decay factor $\alpha \in (0,1)$, assign the following values to a node $v$ in a graph $G$; note that Random Walk Decay considers only walks that enter $v$ for the first time and that, for functional graphs, the denominators vanish:\looseness-1
\begin{equation*}
		\prm^{\alpha}_G(v) = \sum_{(u_1,\dots,u_k,v) \in \Omega(G)} \frac{\alpha^k}{\prod_{i=1}^{k} \deg^+(u_i)}
        \qquad \qquad
        \rwd^{\alpha}_G(v) = \sum_{\substack{(u_1,\dots,u_k,v) \in \Omega(G) \\v \not \in \{u_1,\dots,u_k\}}} \frac{\alpha^k}{\prod_{i=1}^{k} \deg^+(u_i)}
	\end{equation*}


\section{The Power Metric and its Semantics}  \label{sec:expected}

This section provides a conceptual justification for the proposed method of capturing voting power, through presenting equivalent interpretations of its underlying semantics.
The mainstream understanding of power in liquid democracy is currently based on a simple accrual idea \cite{christoff2017binary,golz2021fluid}: the power of an agent $u$ within a delegation graph $G$ amounts to the number of votes they (directly or not) accrue---in our notation $w_u$. We refer to this notion as \emph{nominal weight}. 
Our objective in this paper is to develop an alternative understanding of power which does justice to the fact that delegations may be suspended at any time.

\subsection*{Power as Expected Voting Weight and as Flow of Voting Rights}

We first extend delegation graphs with a parameter $p \in (0,1)$, assigning to each agent $u$ a probability $p$ of delegating (according to an underlying functional delegation graph $G$) and, therefore, a probability $1 - p$ that $u$ suspend their delegation and vote directly. This describes the probability $\P(uv)$ of any given delegation that is captured by an edge $(u,v) \in E$ to be realized,
which in turn can define an $n \times n$ stochastic matrix $[\P(uv)]$ (called {\emph{delegation matrix}}) where each row has value 
$1-p$ on the diagonal entry, and at most one off-diagonal entry with value $p$. Equivalently, the matrix defines a weighted directed graph in the standard way. In what follows we will refer to the delegation matrix $[\P(uv)]$ as $\CalD$. 
Below we present two natural ways in which to interpret $\CalD$:
First, it can be seen as a \emph{lottery on delegation graphs}. Each $\CalD$ is equivalent to a lottery over the set of all functional delegation graphs~$V^V$. Such a lottery is the product probability distribution obtained from all Bernoulli trials given by probabilities $p$, for $u \in V$. The probability of sampling from $\CalD$ an arbitrary functional delegation graph $G = (V,E) \in V^V$ is $\P(G) = \prod_{(u,v) \in V\times V} \P(uv)^{\1_{(u,v) \in E}}(1-\P(uv))^{\1_{(u,v) \notin E}}$. 
    Second, $\CalD$ can be viewed to describe the \emph{flow of voting rights} of a user as the probability with which they transfer to another user via delegations. The delegation matrices are instances of the class of weighted delegation profiles (as studied in the literature of LD \cite{zhang2021power}), as well as of command games \cite{hu2003authoritya,hu2003authorityb} from cooperative game theory.\looseness-1


We explore the implications for the understanding of power with suspendible delegations following the two above interpretations. 
First, we can think of the voting weight of a given user $u$ as a random variable determined by the lottery $\CalD$ defined over all possible delegation graphs. So, $u$'s power in $\CalD$ can be thought of as the expectation on the number of votes $w_u$ that $u$ accrues when the delegation graph is sampled from $\CalD$. This gives rise to a notion of power of a voter that corresponds to their \emph{expected voting weight} (EVW): 
\begin{equation}
    \evw(u) = \E_{G \sim \CalD}[w_u] = \sum_{\D \in V^V} \P(G)\cdot (1 + |\Pred_G(u)|). \label{eq:pow1}
\end{equation}
Formally, for a user $u$, $\evw(u)$ also depends on $p$ and $\CalD$; however, we suppress this dependence in the notation, as it is usually clear from the context and omitting it reduces clutter.
We illustrate \cref{eq:pow1} further via two extreme examples. A node $u$ receiving two direct delegations attains power $1+2p$, whereas a node $v$ at the end of a length-2 path has smaller power, specifically $1+p+p^2,$ due to indirect influence.
In general, if u receives only direct delegations (a star with $n-1$ leaves), Equation \eqref{eq:pow1} simplifies to:
$
   \E[w_u]=1+(n-1)p,
$   
i.e., the expected number of votes cast grows linearly with the number of incoming direct delegations.   
It is worth noticing that this is a special case of the notion of power in weighted delegation graphs studied in \cite{zhang2022tracking}.


We now explore the second interpretation and will show it leads to an equivalent definition to Equation~\eqref{eq:pow1}. We think of the flow of voting rights in the delegation graph as a stochastic process $(X_k)_{0 \leq k}$ where $X_k$ denotes the state (i.e., agent) to which the voting rights flows over time starting from state/user $X_0$. In other words, we view the flow of power in a suspendible delegation profile as a Markov chain.
We are interested in the probability that the votes of a given user $u$ reach a given user~$v$ via the delegation path (if any) that connects the two. This amounts to determining the so-called first-passage probabilities with which voting rights of each user reach any other user. 
The probability that $u$'s voting rights flows to $v$ via the shortest path $(u=u_0,u_1, \ldots,u_{\ell} = v)$ is $p(uv) = \prod_{0 \leq k \leq \ell-1} p$,
if~$u_{i-1}$ delegates to $u_{i}$ for each $i \in [\ell]$, and zero, otherwise.
We can then think of the power of a user $u$ as their own voting rights plus the sum of the first-passage probabilities for any vote from any other user $v$ to $u$, that is:
$    \f(u) = 1+ \sum_{v \in V} p(vu).
$
The notion of power based on first-passage probability is equivalent to the one based on expected voting weight (Equation \eqref{eq:pow1}). 
Intuitively, the voting weight that a user should expect based on the delegation paths made possible by a suspendible delegation profile is equivalent to the probability in which voting rights flow to the user from any other user in the profile.
\begin{theorem} \label{th:equiv}
    For any $\CalD$ derived from a functional delegation graph $G=(V,E)$ and user $u \in V$:
        $\evw(u) = \f(u)$.
\end{theorem}

The resulting notion of power yields the following properties.
For any $u \in \N$, $\lim_{p \to 1} \E_{\D \sim \CalD}[w_u] = w_u(\D)$, that is, as the probability of delegating tends to $1$, $\evw(u)$ converges to $w_u$.  
We also have that $\lim_{p \to 0} \E_{\D \sim \CalD}[w_u] = 1$, that is, as the probability of delegating tends to $0$, $\evw(u)$ converges to $1$, and, hence the system approaches direct democracy. 
Furthermore, for any $u \in \N$, any delegation path of length $\ell$ to $u$ contributes
$        \sum_{1 \leq k \leq \ell} p^k = p \frac{ (1 - p^\ell)}{1 - p}
$    to $\E_{G \sim \CalD}[w_u]$, and as $\ell \to \infty$ the contribution of the path to $u$'s expected weight tends to $\frac{p}{1-p}$ and is therefore upper-bounded by $1$ when $p \leq 0.5$.
It follows that the expected weight is always limited, regardless of the path length; this stands in contrast to nominal weight, where a delegation path of length $\ell$ contributes exactly $\ell$, growing without bound as the chain lengthens. Moreover, the rate of growth of expected weight decreases as the length of the delegation paths increases, unlike in the case of nominal weight, and can never yield more than one extra vote per path in expectation when there exists at least a $50\%$ chance that agents vote directly. 
The presented framework illustrates that, once the suspendibility of delegations is incorporated into liquid democracy, the transitivity of delegations acts as a natural mechanism for limiting power, rather than causing the unconstrained accumulation of influence that liquid democracy has often been criticized for (e.g., \cite{berinsky2025tracking,kahng2021liquid}).


\subsection*{Power as Random-Walk-Based Influence within Delegation Networks}

We now return to the centrality measures introduced in \Cref{sec:prelims}. These yield a third interpretation of our power in the model of suspendible delegations, equivalent to the two discussed above and captured by the Random Walk Decay index.
Both PageRank and Random Walk Decay can be described through the following \emph{random surfing process on a network}. Consider a random surfer that originates from a random node and traverses the network in a random fashion. At each node, the surfer, with probability $p$, chooses one of the outgoing edges uniformly at random, and with probability $(1-p)$ stops surfing. It can be shown that PageRank is proportional to the expected number of visits of the surfer to a given node. 
In contrast, Random Walk Decay counts only the first visit of the random walk to each node rather than all subsequent visits. Hence, it is proportional to the probability that the node is visited at all (see \cite{was2019random} for more details).
In the context of delegative democracy,
given a delegation graph,
$p$ can be interpreted as the probability that a voter does not participate and instead relies on their default delegation, effectively passing their voting power along an outgoing edge. 
Random Walk Decay is better suited for measuring power in this context. We formally justify this in \Cref{sec:centrality_for_power}; intuitively, however, assigning voting power to a voter who has already delegated their vote (i.e., appeared earlier on the path) is counterintuitive.

Both in the model of suspendible delegations captured through a delegation matrix and in the random surfer model, the influence of nodes/agents propagates along network paths/delegation relationships 
according to the probability that a vote/visit reaches them. This analogy relates random walk centrality measures to voting power expressed as the expected voting weight and its flow in delegation-based systems, as discussed above.
Let $\mathcal{G}(\CalD) = (V,E)$ be a directed graph
representing the delegation matrix $\CalD,$
which has an edge for each potential delegation, i.e.,
$E = \{(u,v) \in V^2 : u\neq v \land \P(uv) > 0\}$.

\begin{theorem}
\label{thm:rwd:equivalence}
    For any $\CalD$ derived from a functional graph $G=(V,E)$
    and user $u \in V$:
    $\evw(u) = \f(u) = \rwd^p_{\mathcal{G}(\CalD)}(u)$.
\end{theorem}

\begin{figure}
\centering
\hspace{-2.6cm}
\begin{tikzpicture}
 
\def\s{1.15}

  \node[efc-node] (A1) at (3.2,  1.4) {\efcSL{1.00}{1}};
  \node[efc-node] (A2) at (2.8,  0.6) {\efcSL{1.00}{1}};
  \node[efc-node] (A3) at (2.0,  0.2) {\efcSL{1.00}{1}};
  \node[efc-node] (A4) at (2.0,  1.4) {\efcSL{3.47}{9}};
  \node[efc-node] (A5) at (0.8,  0.2) {\efcSL{3.37}{9}};
  \node[efc-node] (A6) at (-0.4,  1.4) {\efcSL{2.56}{9}};
  \node[efc-node] (A7) at (-0.4, 0.2) {\efcSL{1.75}{3}};
  \node[efc-node] (A8) at (-1.6, 0.2) {\efcSL{1.50}{2}};
  \node[efc-node] (A9) at (-2.8, 0.2) {\efcSL{1.00}{1}};
 
  \draw[efc-edge] (A1) -- (A4);
  \draw[efc-edge] (A2) -- (A4);
  \draw[efc-edge] (A3) -- (A4);
  \draw[efc-edge] (A4) -- (A5);
  \draw[efc-edge] (A5) -- (A6);
  \draw[efc-edge] (A6) -- (A4);
  \draw[efc-edge] (A7) -- (A5);
  \draw[efc-edge] (A8) -- (A7);
  \draw[efc-edge] (A9) -- (A8);

  \node[efc-node] (B9)  at (4.2,  0.2) {\efcSL{3.00}{9}};
  \node[efc-node] (B10) at (5.4,  0.2) {\efcSL{1.75}{3}};
  \node[efc-node] (B11) at (6.6,  0.2) {\efcSL{1.50}{2}};
  \node[efc-node] (B12) at (7.8,  0.2) {\efcSL{1.00}{1}};

  \node[efc-node] (B8)  at (4.2,  1.4) {\efcSL{2.25}{5}};
  \node[efc-node] (B7)  at (5.4,  1.4) {\efcSL{2.50}{4}};
 

  \node[efc-node] (B4)  at (6.6,  1.8) {\efcSL{1.00}{1}};
  \node[efc-node] (B5)  at (7.8,  1.4) {\efcSL{1.00}{1}};
  \node[efc-node] (B6)  at (6.6,  1.0) {\efcSL{1.00}{1}};  
 
  \draw[efc-edge] (B10) -- (B9);
  \draw[efc-edge] (B11) -- (B10);
  \draw[efc-edge] (B12) -- (B11);
  \draw[efc-edge] (B8)  -- (B9);
  \draw[efc-edge] (B7)  -- (B8);
  \draw[efc-edge] (B4)  -- (B7);
  \draw[efc-edge] (B5)  -- (B7);
  \draw[efc-edge] (B6)  -- (B7);
 
\end{tikzpicture}
\caption{A functional graph with two connected components. For each node, we display its power under our conception of influence within the model of suspendible delegations (\textcolor{BrickRed}{red}/top) and its nominal weight (bottom), using $\alpha=0.5$ for the former. }
\label{fig:nominal_vs_expected}
\vspace{-0.3cm}
\end{figure}
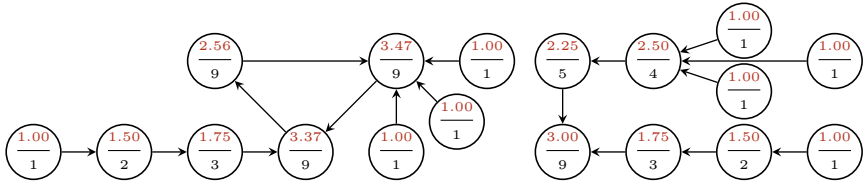

We propose to use voters' expected voting weight ($\evw$), or, nodes' Random Walk Decay centrality ($\rwd$) as the measure of power for the model of suspendible delegations in LD. It is not only more convenient to reason about and handle analytically compared to the existing notions of power for delegative voting systems \cite{colley23measuring,zhang2021power}, but it is also computationally tractable \cite{was2019random}. In particular, for a graph~$G,$ $\rwd^\alpha_G(v)$ corresponds to the PageRank of $v$ in $G-E^+(v)$, which can be efficiently computed~\cite{arasu2002pagerank,das2013fast}.
The metric is illustrated in \Cref{fig:nominal_vs_expected}, alongside nominal weight.
The figure illustrates how influence with respect to our proposed metric fades with delegation distance, how direct delegations dominate indirect ones, and how the resulting power distribution becomes less concentrated.\looseness-1

\section{Axiomatic Analysis of the Proposed Power Metric}
\label{sec:centrality_for_power}

In this section, we articulate the main technical underpinning in favor of $\evw$. We begin (\cref{sec:decayVSpr}) by arguing---through a set of domain-specific axioms for evaluating how power metrics handle delegation cycles---that it is better suited as a measure of influence than PageRank, which has already received significant attention in the relevant literature, as noted in \cref{sec:intro}.
Then (\cref{sec:charact}) we move beyond this comparison to show that $\evw$ is uniquely characterized among all power measures by a set of natural and desirable---in the context of delegative democracy---axioms. 

\subsection{Axiomatic Comparison Against PageRank}
\label{sec:decayVSpr}

Most literature on liquid democracy treats cycles as undesirable, since they are believed to potentially increase abstentions and consequently reduce the voting power of participants and the legitimacy of the process. In acyclic functional networks, the values of PageRank and Random Walk Decay coincide. Thus, there is arguably no reason to prefer one over the other there.
The presence of cycles causes the discussed centrality metrics to diverge, raising the question of which measure is most appropriate then. 
Let us first argue on the relevance of cycles in delegative voting systems, under the model of suspendible delegations.
First, it has been observed \cite{behrens2014principles} that agents intentionally form cycles in real delegative systems. Second, this behavior is explained in theory by the incentive analysis for the model of suspendible delegations executed by \citet{brillcycles}. At a high level, since default delegations expressed by a voter activate only when they do not participate, a group of like-minded participants may intentionally form such a cycle so that their voting weight is still taken into account as long as at least one of them decides to vote.\looseness-1

Our first axiom focuses on two simple yet canonical subclasses of functional graphs: cycles and stars. Given a delegation graph that is a star, the voter corresponding to the central node, when deciding to vote, receives direct support from all voters who do not cast a ballot. In contrast, a node in a cycle receives indirect support of varying levels from its predecessors in the case that the corresponding voter decides to vote while the remaining members of the cycle delegate their ballots. Since direct support should count more than the indirect one, the power of the center of a star should be at least as large as the power of a member of a cycle, provided that the two have the same number of predecessors.\looseness-1

\smallskip
\begin{description}\item[Inverse Distance Monotonicity]
    Consider a cycle on $n\geq 2$ nodes $C_n$ and a star with $n-1\geq 1$ leaves; we call $c_n$ its central node. For each node $v\in C_n$ we want $p(v)\leq p(c_n)$.
\end{description}

\medskip
The second axiom focuses exclusively on cycles and requires that the power of their members is proportional to the size of the cycle. In particular, if two groups of voters form cycles, the power of each member should reflect the size of the corresponding group. This axiom is justified by the fact that members of larger cycles receive the same support as those in smaller cycles at short distances, but continue to accumulate additional contributions beyond the length of the smaller cycle.

\smallskip

\begin{description}\item[Cycle Size Monotonicity]
    Consider two cycles, one on $n$ and one on $n'>n$ nodes, namely $C_n$ and $C_{n'}$. For each pair of nodes $v\in C_n, v'\in C_{n'}$ we want $p(v)< p(v')$. 
\end{description}

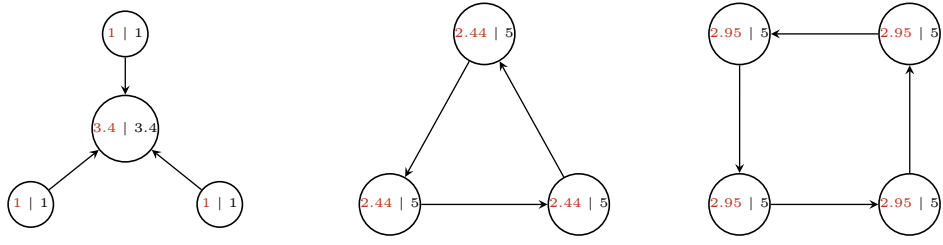
\begin{figure}
\centering
\hspace{-0.3cm}
\begin{tikzpicture}[scale=1.25]
 
\tikzset{
  mysmallnode/.style={
    circle,
    draw=black,
    line width=0.6pt,
    minimum size=0.6cm,
    inner sep=0pt,
    fill=white,
    font=\fontsize{5}{5.5}\selectfont
  },
  mysmalledge/.style={->, draw=black, line width=0.5pt, >=stealth}
}
 
\newcommand{\VSL}[2]{%
  \renewcommand{\arraystretch}{1}%
  \begin{tabular}{@{}c@{}}
    \textcolor{BrickRed}{#1} | #2
  \end{tabular}%
}
 
\node[mysmallnode] (S0) at (0.000,  0.200) {\VSL{3.4}{3.4}};
\node[mysmallnode] (S1) at (0.000,  1.200) {\VSL{1}{1}};
\node[mysmallnode] (S2) at (-1.0,-0.600) {\VSL{1}{1}};
\node[mysmallnode] (S3) at ( 1.0,-0.600) {\VSL{1}{1}};
 
\draw[mysmalledge] (S1) -- (S0);
\draw[mysmalledge] (S2) -- (S0);
\draw[mysmalledge] (S3) -- (S0);
 
\node[mysmallnode] (C1) at (3.800,  1.200) {\VSL{2.44}{5}};
\node[mysmallnode] (C2) at (2.8, -0.60) {\VSL{2.44}{5}};
\node[mysmallnode] (C3) at (4.8, -0.60) {\VSL{2.44}{5}};
 
\draw[mysmalledge] (C1) -- (C2);
\draw[mysmalledge] (C2) -- (C3);
\draw[mysmalledge] (C3) -- (C1);
 
\node[mysmallnode] (D1) at (8.3,  1.2) {\VSL{2.95}{5}};
\node[mysmallnode] (D2) at (6.5,  1.2) {\VSL{2.95}{5}};
\node[mysmallnode] (D3) at (6.5, -0.6) {\VSL{2.95}{5}};
\node[mysmallnode] (D4) at (8.3, -0.6) {\VSL{2.95}{5}};
 
\draw[mysmalledge] (D1) -- (D2);
\draw[mysmalledge] (D2) -- (D3);
\draw[mysmalledge] (D3) -- (D4);
\draw[mysmalledge] (D4) -- (D1);
 
\end{tikzpicture}
\caption{A graph with three connected components illustrating the differences between Random Walk Decay (\textcolor{BrickRed}{red}/left) and PageRank (right) with respect to the Inverse Distance Monotonicity and Cycle Size Monotonicity axioms. Under PageRank, unlike Random Walk Decay, the center of the star has lower power than the cycle nodes despite receiving direct support, while the power of cycle nodes remains unaffected by the size of the cycle.}
\label{fig:axioms}
\vspace{-0.3cm}
\end{figure}

\medskip
Despite their simplicity and relevance in the context of delegative democracy, we show that these two axioms are not satisfied by PageRank. This is due to the fact that, under PageRank, cycles can cause power to increase abruptly for their members and can be seen as a drawback when using it to capture power. In contrast, it is easy to prove that Random Walk Decay centrality does not suffer from this issue. We present an illustration in \Cref{fig:axioms} and the formal theorem follows.

\begin{theorem}
\label{thm:axioms}
Inverse Distance Monotonicity and Cycle Size Monotonicity are satisfied by $\rwd$, but not by $\prm$.
\end{theorem}

We view \cref{thm:axioms} as capturing concrete reasons why, for delegative voting platforms, Random Walk Decay centrality is more suitable than PageRank, despite the latter having already received considerable attention in the relevant literature, while Random Walk Decay has received none.

\subsection{Axiomatic Characterization within Power Metrics}
\label{sec:charact}
We provide a characterization of the proposed measure, establishing it as not only a computationally tractable but also an axiomatically sound alternative to the traditional, more cumbersome game-theoretic indices that have been proposed for measuring voters' power in delegative settings \cite{colley23measuring,zhang2021power}.

\medskip
For the first axiom, let $u$ and $v$ be two endpoints, and $w$ be a voter delegating to $u$. The axiom says that if instead $w$ delegates to $v$, then the sum of powers of $u$ and $v$ stays the same, while the power of any other voter is not affected.
It captures the idea that redirecting a delegation to an endpoint should only redistribute power locally between the affected nodes, with no spillover effects.\looseness-1
\begin{description}
    \item[Endpoint Reassignment] For every graph $G=(V,E)$, endpoints $u,v\in V$ and node $w \in V$, such that $(w,u)\in E$, and graph $H = (V, E \setminus \{(w,u)\} \cup \{(w,v)\})$,
    it holds that $p_H(u) + p_H(v) = p_G(u)+p_G(v)$ and  
    $p_H(x)=p_G(x), \forall x\in V\setminus\{u,v\}.$
\end{description}
\medskip
The next axiom, says that delegating your vote to someone means they effectively gain your share of power.
In particular, say that $u$ and $v$ are endpoints and $v$ additionally does not receive any delegations. If $u$ delegates to $v$, the power of $v$ grows proportionally to the power of $u$.
\begin{description}
    \item[Origin of Endpoint Power] There exists $\alpha>0$ such that for every graph $G=(V,E)$, endpoints $u, v\in V$ such that $\Pred(v)=\emptyset$, and graph $H = (V, E \cup \{(u, v)\})$
    it holds that
    $p_H(v)-p_G(v)=\alpha p_G(u).$
    \end{description}
    \medskip
The third axiom is a simple normalization one. It says that in a graph with only one node, the corresponding voter has a power of $1$; this is consistent with their voting weight. 
    \begin{description}
    \item[Power of the Single User] For $G=(\{v\},\emptyset)$ it holds that $p_G(v)=1$.
    \end{description}
    \medskip
Next, we introduce a notion related to strategyproofness, which requires that a voter’s power does not depend on whom they delegate to, but only on who delegates to them.    
\begin{description}
        \item[Lack-of-Self-Impact] For every graph $G=(V,E)$, nodes $u, v\in V$ such that $(u,v) \in E$, and graph $H = (V, E \setminus \{(u,v)\})$ 
    it holds that
    $p_H(u)=p_G(u)$.
\end{description}
\medskip
Finally, we introduce an axiom requiring the power of a voter to only depend on their connected component. 
This axiom says that only voters who could, in principle, delegate (directly or indirectly) to a node matter for the power of this node. 
\begin{description}
    \item[Locality] For every two graphs $G=(V,E),G'=(V',E')$ and node $v\in V,$ creating a graph $H$ as the disjoint union of $G$ and $G'$ satisfies $p_H(v)=p_G(v)$.
\end{description}
\medskip
The following result builds upon the axiomatic characterization by \citet{was2019random}.
There are, however, important technical and conceptual differences. First, since we restrict our setting to functional graphs, some operations that were allowed in the related proof from \cite{was2019random} are not allowed here, as they would lead outside of the class of functional graphs.  Moreover, we do not need the Random Walk Property axiom, which was required in \cite{was2019random}.
Finally, we do not consider node weights, which also allows for less flexibility and changes how Endpoint Reassignment axiom is defined and used---refer to \cref{sec:concl} for a relevant discussion and generalizations.

\begin{theorem}
\label{thm:rwd:characterization}
    For functional delegation graphs, $\rwd$ is the unique power metric that satisfies Endpoint Reassignment, Origin of Endpoint Power, Power of the Single User, Lack-of-Self-Impact and Locality.
\end{theorem}

\section{
Representative Slates of Powerful Participants
}
\label{sec:peerselection}

In this section we put the above theory at work and conceptualize a power diagnostic tool for decision-making platforms supporting delegations, like LiquidFeedback. The tool should ideally help participants to quickly identify the influential participants of the system and do that in a \emph{representative} manner, that is, in a way that proportionally reflects the delegation structure of the system. Concretely, the problem we are interested in, namely the \emph{Peer Selection Problem}, is the following:
We are given a graph whose nodes represent both voters and candidates and the edges represent delegations, with the goal of selecting $k$ nodes (that is, a slate) which best represent how power is concentrated in the system. 
Building on the mechanics of the Random Walk Decay centrality employed earlier as a metric of voting power,
we propose a method for solving this problem, namely \absorbdecay. We study it from the axiomatic and algorithmic perspective.

Given a delegation graph that has evolved within a delegative democracy platform, it is often useful to select a subset of $k$ participants based on this structure. This selection is not necessarily meant to identify the most influential individuals, but rather those who are sufficiently powerful while collectively providing a meaningful and balanced representation of the broader electorate. This set can serve to summarize the overall landscape of opinions for decision-makers, journalists reporting on the platform, or for internal reporting purposes.
Moreover, identifying a core set of representative participants can help track how opinions evolve over time or facilitate constructive dialogue and agreement.
In large-scale platforms, such a subset may also be needed to form discussion panels for events or public consultations, or to support policy-making processes where only a manageable number of voices can be involved.
In this way, once implemented, such an algorithm may further improve the transparency, effectiveness, and versatility of the collective decision-making process.

\paragraph{The Utilitarian Objective and Further Advantages of Random Walk Decay over PageRank} Before turning to fairness considerations, we discuss a reasonable, yet, rather unrepresentative as we will later see, approach: \textit{selecting the $k$ voters with the highest centrality under a given measure}. When combined with PageRank, or Random Walk Decay, we refer to this method as \emph{\toprank}, and \emph{\topdecay}, respectively.
We will later use these methods as baselines against the \absorbdecay, but, first,
in relation to \cref{sec:decayVSpr}, we further strengthen the case for Random Walk Decay over PageRank in delegative democracy settings, now in the peer selection domain. We mainly reiterating two arguments from \cite{was2019random}; a more extensive discussion can be found there.
Unlike Random Walk Decay, PageRank fails to satisfy Lack-of-Self-Impact, a strategyproofness-related property, where strategyproofness has been identified and long studied as a particularly important desideratum in peer selection settings \cite{alon2011sum,holzman2013impartial,olckers2024manipulation}. A second argument concerns diversity: unlike PageRank, Random Walk Decay favors nodes reachable by structurally diverse regions of the network, leading to selections that more faithfully reflect the underlying structure of the system.\looseness-1

\subsection*{\absorbdecay: A Method for Proportional Representation}

The peer selection problem, with the goal of achieving fair representation of the population, has also been studied by \citet{boldi2011viscous}, \citet{papasotiropoulos2026representation}, and \citet{papasotiropoulos2025proportional}.
Notably, all of the mechanisms proposed and analyzed in these works return, so-called, \emph{hierarchical solutions}, meaning that no node can be selected unless all of its successors are also selected. While such selections are reasonable---as a direct successor of a vertex can be viewed as at least as good a choice as the vertex itself---and may be appropriate in certain settings, we argue below that they can nevertheless be counterintuitive when the input graph arises from a delegative democracy platform.
In contrast, we propose a mechanism, \absorbdecay, that does not follow this structure. In particular, when the selection of $k$ members is based on the delegation preferences of the participants, we take the view that selected voters do not need to delegate their votes further (as they already represent themselves), and thus can be seen as “absorbing” the chain of delegations. Under this perspective, if a participant is selected, their direct predecessor adds little if included as well, since the latter was willing to transfer their entire voting power to the already selected vertex.

We introduce the following class of selection methods. 
For an arbitrary centrality measure $F$, where $F_G(u)$ denotes the centrality of node $u$ in $G$, we define the importance of each selected node $u \in S$ as $F_{G - E^+(S)}(u)$, where the graph $G - E^+(S)$ is obtained by removing all outgoing edges of nodes in $S$.
We then evaluate a set of selected nodes $S \subseteq V$ by the least important node in the set, in terms of $F$, i.e.,
$ F_G(S) = \min_{i \in S} F_{G - E^+(S)}(u).$
The intuition is that every selected node should receive sufficiently strong support from nodes that are not already represented by others in $S$.
Accordingly, we propose the mechanism that selects a group of size $k$ maximizing this score, i.e.,
$\max_{S\subseteq V: |S|=k}\min_{i \in S} F_{G - E^+(S)}(u)$. Combined with PageRank, this yields the rule we call \emph{\absorbdecay}.
Recall that $\rwd^\alpha_G(v)$ corresponds to $\prm^\alpha_{G-E^+(v)}(v)$. Therefore \absorbdecay closely follows the idea of Random Walk Decay centrality in that it does not account for the outgoing edges of the selected nodes.
For simplicity, we assume that the decay factor $\alpha$ approaches 1 for the considered rule.

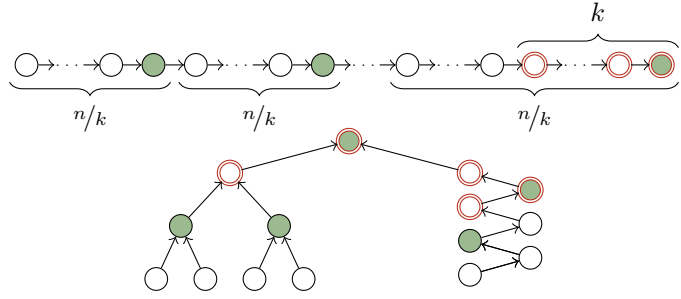
\begin{figure}[t]
  \centering
  \begin{tikzpicture}[scale=0.7]
    \def\y{0}
    \def\x{0.8}
    \node[e4c node] (1) at (1.00*\x, \y) {}; 
    \node[e4c node,draw=none] (2) at (2.00*\x, \y) {$\dots$}; 
    \node[e4c node] (3) at (3.00*\x, \y) {}; 
    \node[e4c node,selected2] (4) at (4.00*\x, \y) {}; 
    
    \node[e4c node] (5) at (5.00*\x, \y) {}; 
    \node[e4c node,draw=none] (6) at (6.00*\x, \y) {$\dots$}; 
    \node[e4c node] (7) at (7.00*\x, \y) {}; 
    \node[e4c node,selected2] (8) at (8.00*\x, \y) {}; 
    \node[e4c node,draw=none] (9) at (9.00*\x, \y) {$\dots$}; 
    
    \node[e4c node] (10) at (10.00*\x, \y) {}; 
    \node[e4c node,draw=none] (11) at (11.00*\x, \y) {$\dots$}; 
    \node[e4c node] (12) at (12.00*\x, \y) {}; 
    \node[e4c node,selected,selected3] (13) at (13.00*\x, \y) {}; 
    \node[e4c node,draw=none] (14) at (14.00*\x, \y) {$\dots$}; 
    \node[e4c node,selected,selected3] (15) at (15.00*\x, \y) {}; 
    \node[e4c node,selected,selected2,selected3] (16) at (16.00*\x, \y) {}; 
    \draw [decorate,decoration={brace,amplitude=5pt,mirror,raise=1.5ex}]
      (0.6*\x,\y) -- (4.4*\x,\y) node[midway,yshift=-20]{$\nicefrac{n}{k}$};
    \draw [decorate,decoration={brace,amplitude=5pt,mirror,raise=1.5ex}]
      (4.6*\x,\y) -- (8.4*\x,\y) node[midway,yshift=-20]{$\nicefrac{n}{k}$};
    \draw [decorate,decoration={brace,amplitude=5pt,mirror,raise=1.5ex}]
      (9.6*\x,\y) -- (16.4*\x,\y) node[midway,yshift=-20]{$\nicefrac{n}{k}$};
      
    \draw [decorate,decoration={brace,amplitude=5pt,raise=1.5ex}]
      (12.6*\x,\y) -- (16.4*\x,\y) node[midway,yshift=20]{$k$};
   
    \path[e4c path]
    (1) edge[e4c edge]  (2)
    (2) edge[e4c edge]  (3)
    (3) edge[e4c edge]  (4)
    (4) edge[e4c edge]  (5)
    (5) edge[e4c edge]  (6)
    (6) edge[e4c edge]  (7)
    (7) edge[e4c edge]  (8)
    (8) edge[e4c edge]  (9)
    (9) edge[e4c edge]  (10)
    (10) edge[e4c edge]  (11)
    (11) edge[e4c edge]  (12)
    (12) edge[e4c edge]  (13)
    (13) edge[e4c edge]  (14)
    (14) edge[e4c edge]  (15)
    (15) edge[e4c edge]  (16)
    ;
  \end{tikzpicture}

\vspace{-0.2cm}
  \begin{tikzpicture}[x=5cm, y=2.8cm]
    \node[e4c node,selected,selected2,selected3] (1)  at (0.51, 1.50) {};
    \node[e4c node,selected,selected3]           (2)  at (0.195, 1.35) {};
    \node[e4c node,selected2,selected3]          (3)  at (0.065, 1.10) {};
    \node[e4c node,selected2]                    (4)  at (0.325, 1.10) {};
    \node[e4c node]                              (5)  at (0.00, 0.85) {};
    \node[e4c node]                              (6)  at (0.13, 0.85) {};
    \node[e4c node]                              (7)  at (0.26, 0.85) {};
    \node[e4c node]                              (8)  at (0.39, 0.85) {};
    \node[e4c node,selected,selected3]           (9)  at (0.83, 1.35) {};
    \node[e4c node,selected,selected2,selected3] (10) at (0.99, 1.27) {};
    \node[e4c node,selected]                     (11) at (0.83, 1.19) {};
    \node[e4c node]                              (12) at (0.99, 1.11) {};
    \node[e4c node,selected2]                    (13) at (0.83, 1.03) {};
    \node[e4c node]                              (14) at (0.99, 0.95) {};
    \node[e4c node]                              (15) at (0.83, 0.87) {};
    \path[e4c path]
      (15) edge[e4c edge] (14)
      (14) edge[e4c edge] (13)
      (2)  edge[e4c edge] (1)
      (3)  edge[e4c edge] (2)
      (4)  edge[e4c edge] (2)
      (5)  edge[e4c edge] (3)
      (6)  edge[e4c edge] (3)
      (7)  edge[e4c edge] (4)
      (8)  edge[e4c edge] (4)
      (9)  edge[e4c edge] (1)
      (10) edge[e4c edge] (9)
      (11) edge[e4c edge] (10)
      (12) edge[e4c edge] (11)
      (13) edge[e4c edge] (12)
      (14) edge[e4c edge] (13)
      (15) edge[e4c edge] (14)
    ;
  \end{tikzpicture}
  \caption{%
    \textbf{Top:} Selecting $k$ nodes from the $n$-path ($n$ divisible by $k$).
    \topdecay, or, more generally, hierarchical solutions (\textcolor{BrickRed}{red} double lines) choose the last $k$ nodes of the path.
    \absorbdecay (\textcolor{OliveGreen}{green} shading) selects nodes evenly splitting the path.
    \textbf{Bottom:} Selecting $k=5$ nodes from a directed in-tree with two unbalanced branches of equal size.
    \topdecay (\textcolor{BrickRed}{red} double lines) selects mostly from the right-hand side.
    \absorbdecay (\textcolor{OliveGreen}{green} shading) splits the tree into subtrees of size $3$.}
  \label{fig:paths-and-tree}
  \vspace{-0.3cm}
\end{figure}

Towards revealing the fair nature of \absorbdecay, we start by illustrating it on functional graphs which, apart from the main considered scenario in LiquidFeedback, also happen to be simple enough to capture what fairness may mean there. In the peer selection setting, they can be viewed as elections with 1-approval ballots and were also the exclusive focus of \citet{boldi2011viscous} and \citet{papasotiropoulos2026representation}, who studied the same problem.  
We begin with an analysis of paths.
Consider one with $n$ vertices (say divisible by $k$). Naturally, the endpoint of the path will be selected, but then, 
if the second-to-last node is selected, the endpoint will have no incoming edges in $G - E^+(S)$, and its centrality will be minimized.  
Therefore, the \absorbdecay rule avoids such a selection.  
Instead, it aims to select nodes in such a way that they have equal support in the graph without their outgoing edges: it selects nodes $\nicefrac{n}{k}, 2 \cdot \nicefrac{n}{k}, \dots, n$. Any hierarchical solution, including \topdecay and \toprank, would pick the last $k$ vertices instead. 
For an illustration see \Cref{fig:paths-and-tree} (Top).
Consider now an arbitrary (connected) functional graph.
Nodes from the cycle, if any, have maximal $\prm$.
If the graph has a cycle, then \absorbdecay would select one node from there, say $v$.
The graph $G - E^+(v)$ is a tree.
There, roughly speaking, \absorbdecay splits this tree into parts of equal size, selecting the root from each. 
\Cref{fig:paths-and-tree} (Bottom) illustrates this idea in a particular instance.
The above discussion highlights not only the conceptual difference of our method from \toprank and \topdecay, but, importantly, also from the existing approaches \cite{boldi2011viscous,papasotiropoulos2026representation,papasotiropoulos2025proportional} which would also make selections in a hierarchical manner. 
We conclude the discussion on functional graphs by presenting a tractability result.

\begin{theorem}
	\label{thm:polynomial}
	Given a functional graph $G$, $\absorbdecay(G,k)$ can be computed in polynomial time.
\end{theorem}

We now go beyond functional graphs. For these, we adopt again the axiomatic approach and we will formally argue in favor of \absorbdecay over \topdecay and \toprank when aiming for proportional representation. In contrast to \topdecay, the method \toprank has been previously considered by both \citet{boldi2011viscous} and \citet{papasotiropoulos2025proportional}. In the latter paper it was shown not to satisfy a fairness axiom called \emph{Clique-Entitlement}. 
Clique-Entitlement builds on the literature on proportionality in multiwinner elections and captures the idea that each node should have an influence over a $\frac{k}{n}$ fraction of the committee. Accordingly, a coherent group that mutually approve each other should share influence evenly among its members and, if sufficiently large, should be adequately represented.\looseness-1

\begin{description}
    \item[Clique-Entitlement] For every graph $G=(V,E)$, if there exists a connected component $S \subseteq V$ that forms a clique, then at least $\lfloor\frac{|S|}{n} \cdot k\rfloor$ members of the selected committee must belong to $S$.
\end{description} 

The following result constitutes the main technical contribution of the section and provides evidence in favor of the fair nature of \absorbdecay, even on general graphs.

\begin{theorem}
\label{thm:cliqueentitlement}
Clique-Entitlement is satisfied by \absorbdecay but not by \toprank or \topdecay.
\end{theorem}

We conclude by revisiting 
\cref{thm:polynomial}, this time by going beyond the case of functional graphs.
The result relates to works on complexity of group-centralities \citep{angriman2021group,chen2016efficient,zhao2014measuring}, making it of independent interest.\looseness-1

\begin{theorem}
	\label{thm:nphard}
	Given an input $(G,k)$, it is NP-hard to compute $\absorbdecay(G,k)$.
\end{theorem}

\section{Conclusions and Discussion}
\label{sec:concl}

This paper proposes and studies a novel power metric for delegation-based collective decision making processes as supported by software like LiquidFeedback. We showed that the proposed metric exhibits intuitive axiomatic properties and supports the development of power analytics tools that can readily be implemented within LiquidFeedback and similar platforms.

We conclude by observing that the proposed power metric extends beyond our core model while retaining its desirable axiomatic properties. First, we note that the original definition of Random Walk Decay applies to arbitrary graphs, i.e., beyond functional, even with weighted nodes and edges~\cite{was2019random}. 
Node weighted graphs are particularly relevant in blockchain-based delegative voting settings, for instance in DAOs \cite{weidener2025delegated}, where users may additionally possess voting weights determined by their stake. Edge weights, on the other hand, could naturally capture ranked delegation settings, which have also received considerable attention in the liquid democracy literature \cite{brill2022liquid,colley2022unravelling,kavitha2022popular,utke2023anonymous}.
An axiomatic characterization as a power metric (in the spirit of \cref{thm:rwd:characterization}) can still be derived for general graphs, this time closely following the result from \cite{was2019random}. However, for general graphs it is somewhat unclear whether all of the required axioms admit an intuitive interpretation in the context of delegative democracy. 
Moreover, voting probabilities may differ across participants; while the interpretation in terms of expected voting weight transfers directly, capturing this formally through Random Walk Decay would require a generalized version of the metric with node-dependent decay values along a path.
Nodes' power under $\rwd$ continues to be computable efficiently for general graphs, and thus so can \topdecay. It remains open to develop a representative-selection mechanism that preserves the properties of \absorbdecay\ while remaining polynomial on general graphs.\looseness-1

\section*{Acknowledgements} 
D. Grossi and A. Nitsche acknowledge support by the European Union under the Horizon Europe project Perycles (Participatory Democracy that Scales). G. Papasotiropoulos and P. Skowron were supported by the European Union (ERC, PRO-DEMOCRATIC, 101076570). Views and opinions expressed are however those of the authors only and do not necessarily reflect those of the European Union or the European Research Council. Neither the European Union nor the granting authority can be held responsible for them.
O. Skibski was supported by the National Science Centre under Grant No. 2023/50/E/ST6/00665.
T. Wąs was partially supported by EPSRC under grant EP/X038548/1.

\vspace{-0.3cm}
\begin{figure}[h]
\includegraphics[width=0.25\linewidth]{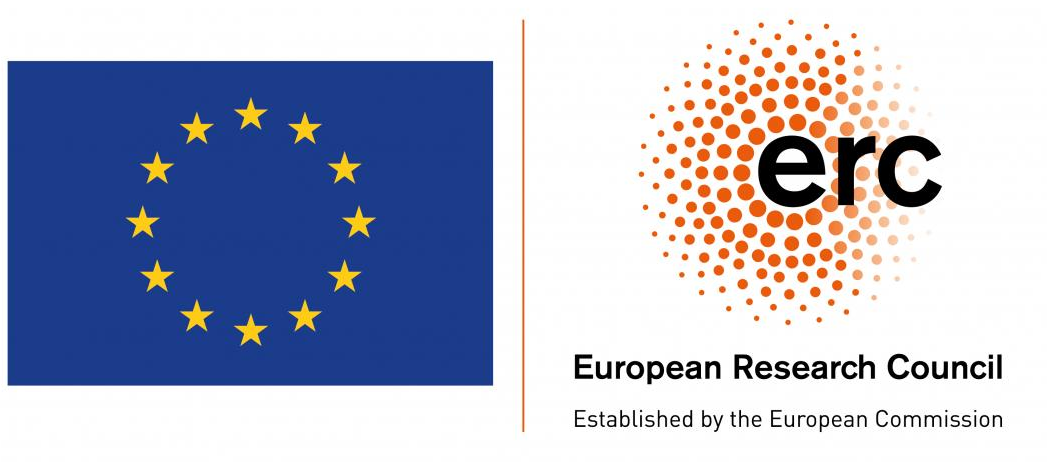}
\hspace{0.05cm}\raisebox{0.65cm}{\includegraphics[width=0.15\linewidth]{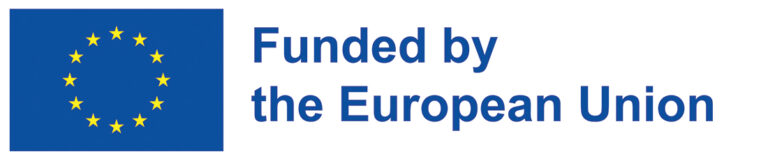}}
\end{figure}

\bibliographystyle{abbrvnat}
\bibliography{sample}
\appendix
\section*{Technical Appendix}

\subsection*{Proof of \cref{th:equiv}}
The claim is proven by the following series of equivalences:
\begin{align*}
    \evw(u) &= \sum_{\D \in \N^\N} \P(\D)\cdot w_u(\D) = \sum_{\D \in \N^\N} \P(\D)\cdot (1 + \sum_{v \in \N} \1_{v \in \Pred(u)})
    \\
    &= \sum_{\D \in \N^\N} \P(\D) \cdot  1 + \sum_{\D \in \N^\N} \P(\D) \cdot \sum_{v \in \N} \1_{v \in \Pred(u)}  = 1 + \sum_{\D \in \N^\N} \P(\D) \cdot \sum_{v \in \N} \1_{v \in \Pred(u)}  = 1 + \sum_{v \in \N} p(vu)  \\
    &= \f(u).
\end{align*}
The second-last equivalence holds for the following reason. We consider the sum over the nominal weight of $i$ in an arbitrary graph $G$, that is, the number of delegation paths linking some user $v$ to $i$, weighted by the probability of $G$ (according to the given suspendible delegation profile $\CalD$). This amounts to the sum over the probability of each such path from $v$ to $u$, which in turn is precisely the sum over all first-passage probabilities from any $v$ to $u$.
\hfill\qed

\subsection*{Proof of \cref{thm:rwd:equivalence}}
Fix delegation matrix $\CalD$ and user $u \in V$,
and denote $\mathcal{G}(\CalD)=(V,E)$.
The fact that $\evw(u) = \f(u)$ is already proved in \Cref{th:equiv}.
Thus, let us focus on showing that $\f(u) = \rwd^p_{\mathcal{G}(\CalD)}(u)$.

First, since in $\CalD$, for each $v \in V$ there is at most one other user $w \in V \setminus \{v\}$
such that $\P(vw) > 0$,
in the graph $\mathcal{G}(\CalD)$ every node has the out-degree of at most one.
This implies that the definition of Random Walk Decay simplifies to
\[
    \rwd^p_{\mathcal{G}(\CalD)}(u) = \sum_{\substack{(v_1,\dots,v_k,u) \in \Omega(G) \\u \not \in \{v_1,\dots,v_k\}}} p^k.
\]

Next, we can show that since in $\mathcal{G}(\CalD)$
every node has the out-degree of at most one,
it holds that every walk $(v_1,\dots,v_k,u) \in \Omega(G)$
such that $u \not \in \{v_1,\dots,v_k\}$
actually is a path, i.e.,
there is no $i \neq j$ such that $v_i = v_j$.
Assume otherwise, and take such $i,j$ with maximum $j$.
Denote $v_{k+1} = u$, and observe that necessarily $v_{i+1} \neq v_{j+1}$
(as either $j < k$, and then we would contradict maximality of $j$,
or if $j=k$, we contradict the fact that $u \not \in \{v_1,\dots,v_k\}$).
However, this means that $v_i=v_j$ has to outgoing edges,
one to $v_{i+1}$ and one to $v_{j+1}$, which contradicts our assumptions.
Therefore, indeed every walk $(v_1,\dots,v_k,u) \in \Omega(G)$
such that $u \not \in \{v_1,\dots,v_k\}$ is a path.
This, in turn implies that 
\[
    \rwd^p_{\mathcal{G}(\CalD)}(u) = 
    \sum_{\substack{(v_1,\dots,v_k,u) \in \Omega(G) \\u \not \in \{v_1,\dots,v_k\}}} p^k
    = \sum_{v \in V} p(vu)
    = \f(u),
\]
which concludes the proof.
\hfill\qed

\subsection*{Proof of \cref{thm:axioms}}
The proof is immediate and follows from straightforward calculations. We start with showing that Cycle Size Monotonicity is not satisfiable by PageRank.
Fix a directed cycle $C_n=(V,E)$, for some integer $n$. Since $C_n$ satisfies $\deg^+(u) = 1$ for all $u \in V$, it holds that
\[
\mathrm{PR}^\alpha_{C_n}(v) = \sum_{(u_1,\dots,u_k,v) \in \Omega(C_n)} \alpha^k,
\]
Fix any node $v \in V$. For each $k \geq 0$, there is exactly one walk of length $k$ ending at $v$ in $C_n$, namely the walk that traces the cycle backwards $k$ steps from $v$. Hence the paths of each length contribute exactly $\alpha^k$, and summing over all $k \geq 0$ gives the geometric series which converges for $|\alpha| < 1$.
\[
\mathrm{PR}^\alpha_{C_n}(v) = \sum_{k=0}^{\infty} \alpha^k = \frac{1}{1-\alpha},
\]
 Since this holds for every $v$ and is independent of $n$, the PageRank of every node in any directed cycle is $\frac{1}{1-\alpha}$, regardless of its size.

We now move to the Random Walk Decay centrality, towards showing that it satisfies Cycle Size Monotonicity. Again, fix some graph that is a cycle $C_n=(V,E)$. The denominator
in the definition of the metric equals $1$ for every path, and the formula reduces to
\[
\mathrm{RWD}^\alpha_{C_n}(v) = \sum_{\substack{(u_1,\dots,u_k,v)\,\in\,\Omega(C_n)\\v\notin\{u_1,\dots,u_k\}}} \alpha^k.
\]
Fix any node $v\in V$. For each $k \geq 0$, there is exactly one walk of length
$k$ ending at $v$ in $C_n$. This walk avoids $v$ in its interior if and only
if it has not yet completed a full traversal of the cycle, i.e.\ $k \leq n-1$.
For $k \geq n$ the unique walk of length $k$ ending at $v$ must pass through
$v$.
Hence the admissible lengths are exactly $k \in \{0, 1, \dots, n-1\}$, giving
\[
\mathrm{RWD}^\alpha_{C_n}(v)
    = \sum_{k=0}^{n-1} \alpha^k
    = \frac{1-\alpha^n}{1-\alpha},
\]
which is strictly increasing in $n,$
for $\alpha \in (0,1)$.

Regarding the Inverse Distance Monotonicity axiom, similarly to before, we can compute the two indices for a star graph, say $S_n$, with $n-1$ leaves:
\[
\mathrm{RWD}^\alpha_{S_n}(c) = \mathrm{PR}^\alpha_{S_n}(c) = 1+(n-1)\alpha.
\]
Then, 
\[
\mathrm{RWD}^\alpha_{S_n}(c) - \mathrm{RWD}^\alpha_{C_n}(v)
= 1 + (n-1)\alpha - \sum_{k=0}^{n-1}\alpha^k
= (n-1)\alpha - \sum_{k=1}^{n-1}\alpha^k
= \sum_{k=1}^{n-1}\alpha - \sum_{k=1}^{n-1}\alpha^k
= \sum_{k=1}^{n-1}(\alpha - \alpha^k) \geq 0.
\]

On the other hand, for PageRank, we have $\mathrm{PR}^\alpha_{C_n}(v) = \frac{1}{1-\alpha}$ and
$\mathrm{PR}^\alpha_{S_n}(c) = 1+(n-1)\alpha$, and any node from the cycle achieves a higher centrality score if
\[
\frac{1}{1-\alpha} > 1+(n-1)\alpha
\iff \alpha > \frac{n-2}{n-1},
\]
which concludes the proof.
\hfill \qed

\subsection*{Proof of \cref{thm:rwd:characterization}}
The proof consists of two parts:
first we will show that Random Walk Decay indeed satisfies all of the axioms,
which will be followed by a proof
that an arbitrary centrality measure satisfying all of the axioms
is equal to Random Walk Decay.

Let us begin by showing that Random Walk Decay indeed satisfies all of the axioms.
Observe that since we consider graphs in which out-degree of each node is at most one,
the definition of Random Walk Decay in such graphs simplifies to
\[
\rwd^{\alpha}_{G}(x) =
    \sum_{\substack{(u_1,\dots,u_k,x) \in \Omega(G) \\x \not \in \{u_1,\dots,u_k\}}} \alpha^k.
\]
Let us consider the axioms one by one.

For Endpoint Reassignment, 
consider an arbitrary graph $G=(V,E)$,
endpoints $u,v\in V$ and node $w \in V$
such that $(w,u)\in E$,
and let $H = (V, E \setminus \{(w,u)\} \cup \{(w,v)\})$.
Fix an arbitrary node $x \in V\setminus\{u,v\}.$
Since $u$ and $v$ are endpoints
no walk that finishes at $x$ can visit $u$ or $v$ before,
thus the change does not affect such walks, i.e.,
$\{ (u_1,\dots,u_k,x) \in \Omega(G) : x \not \in \{u_1,\dots,u_k\} \} = 
\{ (u_1,\dots,u_k,x) \in \Omega(H) : x \not \in \{u_1,\dots,u_k\} \}$
Hence,
\[
    \rwd^{\alpha}_{G}(x) =
    \sum_{\substack{(u_1,\dots,u_k,x) \in \Omega(G) \\x \not \in \{u_1,\dots,u_k\}}} \alpha^k
    = 
    \sum_{\substack{(u_1,\dots,u_k,x) \in \Omega(H) \\x \not \in \{u_1,\dots,u_k\}}} \alpha^k
    =
    \rwd^{\alpha}_{H}(x).
\]
Moreover, every walk that ended in either $u$ or $v$
corresponds to exactly one walk that still ends in either $u$ or $v$.
Thus,
\begin{multline*}
    \rwd^{\alpha}_{G}(u) + \rwd^{\alpha}_{G}(v) =
    \sum_{\substack{(u_1,\dots,u_k,x) \in \Omega(G) \\x \in \{u,v\}, \ x \not \in \{u_1,\dots,u_k\}}} \alpha^k = 
    \sum_{\substack{(u_1,\dots,u_k,x) \in \Omega(H) \\x \in \{u,v\}, \ x \not \in \{u_1,\dots,u_k\}}} \alpha^k
    = \rwd^{\alpha}_{H}(u) + \rwd^{\alpha}_{H}(v).
\end{multline*}
Therefore, Endpoint Reassignment is satisfied by Random Walk Decay.

For Lack-of-Self-Impact,
consider an arbitrary graph $G=(V,E)$,
nodes $u, v\in V$ such that $(u,v) \in E$,
and let $H = (V, E \setminus \{(u,v)\})$
Observe that every walk that finishes in $u$
and does not visit $u$ before,
cannot traverse through the edge we delete from $G$,
hence it is unaffected by its deletion.
Formally,
$\{ (u_1,\dots,u_k,u) \in \Omega(G) : u \not \in \{u_1,\dots,u_k\} \} = 
\{ (u_1,\dots,u_k,u) \in \Omega(H) : u \not \in \{u_1,\dots,u_k\} \}$
Thus,
\[
    \rwd^{\alpha}_{G}(u) =
    \sum_{\substack{(u_1,\dots,u_k,u) \in \Omega(G) \\u \not \in \{u_1,\dots,u_k\}}} \alpha^k
    = 
    \sum_{\substack{(u_1,\dots,u_k,u) \in \Omega(H) \\u \not \in \{u_1,\dots,u_k\}}} \alpha^k
    =
    \rwd^{\alpha}_{H}(u).
\]
Therefore, Lack-of-Self-Impact is satisfied by Random Walk Decay.

For Origin of Endpoint Power,
consider an arbitrary graph $G=(V,E)$ and 
endpoints $u, v\in V$ such that $\Pred(v)=\emptyset$, 
and let $H = (V, E \cup \{(u, v)\})$.
Observe that every walk that ends in node $v$ in graph $H$,
except for a length-zero walk, $(v)$, passes through node $u$
just before reaching $v$.
Thus,
\begin{align*}
    \rwd^{\alpha}_{H}(v) &= \sum_{\substack{(u_1,\dots,u_k,v) \in \Omega(H) \\v \not \in \{u_1,\dots,u_k\}}} \alpha^k = 1 +
    \sum_{\substack{(u_1,\dots,u_{k-1},u,v) \in \Omega(H) \\v \not \in \{u_1,\dots,u_{k-1},u\}}} \alpha^k =
    1 + \alpha \cdot 
    \sum_{\substack{(u_1,\dots,u_{k-1},u) \in \Omega(H) \\v \not \in \{u_1,\dots,u_{k-1},u\}}} \alpha^{k-1}. 
\end{align*}
Since $v$ is an endpoint, $v$ and $u$ are not part of any cycle in $H$.
Thus, it is not possible that a walk that ends in $u$
visits $v$ or $u$ before, which means that
\(
    \{(u_1,\dots,u_{k-1},u) \in \Omega(H) :v \not \in \{u_1,\dots,u_{k-1},u\}\} =
    \{(u_1,\dots,u_{k-1},u) \in \Omega(H)\} =
    \{(u_1,\dots,u_{k-1},u) \in \Omega(H) :u \not \in \{u_1,\dots,u_{k-1},u\}\}.
\)
This further implies that
\[
    \sum_{\substack{(u_1,\dots,u_{k-1},u) \in \Omega(H) \\v \not \in \{u_1,\dots,u_{k-1},u\}}} \alpha^{k-1} = \rwd^\alpha_H(u) = \rwd^\alpha_G(u),
\]
where the second equality comes from the fact that Random Walk Decay satisfies Lack-of-Self-Impact.
This, together with the fact that
$    \rwd^\alpha_G(v) = 1
$
and the previous equality, yields that
\[
    \rwd^{\alpha}_{H}(v) =
    1 + \alpha \cdot 
    \sum_{\substack{(u_1,\dots,u_{k-1},u) \in \Omega(H) \\v \not \in \{u_1,\dots,u_{k-1},u\}}} \alpha^{k-1}
    = \rwd^\alpha_G(v) + \alpha \cdot \rwd^\alpha_G(u).
\]
Therefore, Origin of Endpoint Power is satisfied by Random Walk Decay.

For Power of the Single User, observe that in a single-node graph $G=(\{v\},\emptyset)$ there is only a single trivial walk $(v) \in \Omega(G)$.
Thus, by the definition
\[
    \rwd^{\alpha}_{(\{v\},\emptyset)}(v) = \frac{\alpha^0}{1} = 1.
\]

Finally, for Locality, 
consider arbitrary graphs $G=(V,E),G'=(V',E')$ and node $v \in V$,
and let $H$ be the disjoint union of $G$ and $G'$.
Observe that the set of walks that finish in $v$
is identical in $G$ and $H$.
Thus,
$\{ (u_1,\dots,u_k,v) \in \Omega(G) : v \not \in \{u_1,\dots,u_k\} \} = 
\{ (u_1,\dots,u_k,v) \in \Omega(H) : v \not \in \{u_1,\dots,u_k\} \}$,
and
\[
    \rwd^{\alpha}_{G}(v) =
    \sum_{\substack{(u_1,\dots,u_k,v) \in \Omega(G) \\v \not \in \{u_1,\dots,u_k\}}} \alpha^k
    = 
    \sum_{\substack{(u_1,\dots,u_k,v) \in \Omega(H) \\v \not \in \{u_1,\dots,u_k\}}} \alpha^k
    =
    \rwd^{\alpha}_{H}(v).
\]
Therefore, Locality is satisfied by Random Walk Decay.

In the reminder of the proof
let us consider and arbitrary centrality measure $F$
that satisfies all of our axioms.
We will prove that for every graph $G=(V,E)$
and node $v \in V$
it holds that $F_G(v)=\rwd^\alpha_G(v)$,
where $\alpha$ is given from Origin of Endpoint Power.

The proof follows an induction on the number of edges
in the connected component that contains $v$ in $G$.
For the basis of induction,
fix graph $G=(V,E)$ and node $v \in V$, and
assume that $v$ belongs to a connected component with zero edges.
In particular, this means that it does not have any incoming or outgoing edges.
In such a case, by $G' = (V \setminus \{v\}, E)$ let us denote graph $G$ with $v$ removed.
Observe that $G$ can be seen as a disjoint union of $G'$ and $(\{v\},\emptyset)$.
Thus, by Locality and Power of the Single User
we obtain that
\[
    F_G(v) = F_{(\{v\},\emptyset)}(v) = 1 = \rwd^\alpha_G(v).
\]
This concludes the proof of the basis of induction.

Now, assume that we know that
\(
    F_G(v) = \rwd^\alpha_G(v)
\)
for every graph $G=(V,E)$ and node $v \in V$
such that the connected component to which $v$ belongs
has at most $m$ edges, for certain $m \in \mathbb{N}$.
Then, consider an arbitrary graph $G=(V,E)$ and node $v \in V$
such that the connected component to which $v$ belongs
has $m + 1$ edges.
We will prove that it still holds that
\(
    F_G(v) = \rwd^\alpha_G(v)
\)
by considering three cases based on the
number of incoming and outgoing edges of $v$.

\textbf{\underline{Case 1:}} $v$ has an outgoing edge.

First, let us assume that $v$ has an outgoing edge in $G$,
i.e., there exists $(v,u) \in E$, for some $u \in V$.
Then, consider graph with edge $(v,u)$ removed, i.e.,
$G'=(V,E \setminus \{(v,u)\})$.
Observe that in $G'$,
the number of edges in the connected component to which $v$ belongs
has at most $m$ edges.
Hence, from the inductive assumption, we get
\(
    F_{G'}(v) = \rwd^\alpha_H(v).
\)
Then, from the fact that Lack-of-Self-Impact
is satisfied by both $F$ and Random Walk Decay, we get that
\[
    F_G(v) = F_{G'}(v) = \rwd^\alpha_{G'}(v) = \rwd^\alpha_G(v).
\]
This concludes the analysis of this case.

\textbf{\underline{Case 2:}} $v$ is an endpoint with more than one incoming edge.

Second, let us assume that $v$ is an endpoint and has at least two incoming edges, i.e.,
there exist $(u,v), (u',v) \in E$ for some $u,u' \in V \setminus \{v\}$.
Now, let us add an arbitrary isolated node $w \not \in V$ to graph $G$.
Formally, let $G'$ be a disjoint union of graph $G$ and $(\{w\},\emptyset)$, i.e.,
$G' = (V \cup \{w\},E)$.
Observe that by Locality and Power the Single User, we get that
\[
    F_{G'}(v) = F_G(v) \quad \mbox{and} \quad F_{G'}(w) = F_{(\{w\},\emptyset)}(w) = 1.
\]

Now, consider the graph $G''$ obtained from $G'$ by reassigning
the outgoing edge of node $u'$ to node $w$ instead of $v$.
Formally, let $G'' = (V \cup \{w\},E \setminus \{(u',v)\} \cup \{(u',w)\})$.
Since both $v$ and $w$ are endpoints,
by Endpoint Reassignment we get that
\[
    F_{G''}(v) + F_{G''}(w) = F_{G'}(v) + F_{G'}(w) = F_G(v) + 1.
\]

On the other hand, in graph $G''$ both $v$ and $w$ are in connected components
that have at most $m$ edges 
(as each edge present in one of these components,
was present in the connected component of $v$ in $G$ that had $m+1$ edges,
and each of the components of $v$ and $w$ in $G''$ contains at least one edge).
Thus, from the inductive assumption, we know that
\[
    F_{G''}(v) = \rwd^\alpha_{G''}(v) 
    \quad \mbox{and} \quad 
    F_{G''}(w) = \rwd^\alpha_{G''}(w).
\]
Combining it all together, we obtain
\[
    F_G(v) = \rwd^\alpha_{G''}(v) + \rwd^\alpha_{G''}(w) - 1 = \rwd^\alpha_G(v),
\]
where the second equality comes from the fact that
Random Walk Decay also satisfies our axioms,
so an analogous equations as for $F$ hold for it as well.
This concludes the analysis of these case.

\textbf{\underline{Case 3:}} $v$ is an endpoint with a single incoming edge.

Finally, let us assume that $v$ is an endpoint
with a single incoming edge $(u,v)\in E$
for some node $u \in V$.
Consider graph $G'$ obtained from $G$ be removing edge $(u,v)$, i.e.,
$G' = (V, E \setminus \{(u,v)\})$.
Since $u$ has only one outgoing edge,
it is an endpoint in $G'$,
and so is $v$.
Thus, from Origin of Endpoint Power
we know that
\[
    F_{G}(v) - F_{G'}(v) = \alpha F_{G'}(u).
\]

Observe that in graph $G'$ the component to which $u$ belongs contains exactly $m$ edges,
and the component to which $v$ belongs---zero.
Hence, from the inductive assumption we get that
\[
    F_{G'}(u) = \rwd^\alpha_{G'}(u)
    \quad \mbox{and} \quad
    F_{G'}(v) = \rwd^\alpha_{G'}(v).
\]
Putting it altogether, we get that
\[
    F_{G}(v) = \alpha F_{G'}(u) + F_{G'}(v)
    = \alpha \rwd^\alpha_{G'}(u) + \rwd^\alpha_{G'}(v)
    = \rwd^\alpha_G(v),
\]
where the last equality comes from the fact that
Random Walk Decay also satisfies Origin of Endpoint Power.
This concludes the analysis of this case.

Since the three cases we have considered were exhaustive,
by the power of induction,
we know that $F_G(v) = \rwd^\alpha_G(v)$
for every graph $G=(V,E)$ and node $v \in V$.\hfill \qed

\subsection*{Proof of \cref{thm:polynomial}}

It holds that if $G = (V,E)$ is a functional graph
and $S \subseteq V$ is an arbitrary subset of nodes,
then in the graph $G - E^+(S)$ each node $i \in S$
is a root of some in-tree,
i.e., a connected functional graph without a cycle.
This means that the Random Walk Decay centrality of $i$ for $\alpha \rightarrow 1$
is just equal its PageRank and hence to the number of $i$'s predecessors in this graph.
Hence,
$\rwd^{\alpha \rightarrow 1}_G(S) = \min_{i \in S} |\Pred_{G - E^+(S)}(i)|.$
Therefore, we get that
the method chooses a subset of nodes $S \subseteq V$
that maximizes the minimal number of predecessors of nodes from $S$
in $G - E^+(S)$.
To find such a subset,
given parameters $p,\ell \in \mathbb{N}$ and a functional graph $G = (V,E)$,
we ask whether there is a subset $S \subseteq V$ of size $\ell$,
such that each node $i \in S$ has at least $p$ predecessors in $G - E^+(S)$.
We will denote this decision problem by $\Pi(G,p,\ell).$
Deciding $\Pi(G,p,\ell)$ for all possible values of $p$ (which are upper bounded by $|V|$) and selecting the set $S$ that witnesses that $\Pi(G,p,k)$ is a \textit{yes}-instance for the maximum possible value of $p,$ solves the optimization version of \absorbdecay.
  
It holds that $\Pi(G,p,\ell)$ can be solved in polynomial time for in-trees~\citep{frederickson1991optimal,perl1981max}.
We refer to such a procedure as $\texttt{alg-tree}(G,p,\ell)$.
In what follows, we extend $\texttt{alg-tree}(G,p,\ell)$
first to arbitrary connected functional graphs (possibly with a cycle),
and then to all functional graphs (possibly disconnected).

Suppose that $G$ is an arbitrary connected functional graph. 
Observe that $G$ can have at most one cycle
and we can assume that at least one node will be selected form the cycle
(if no node is selected to $S$ from the cycle,
then we can take a selected node that is closest to the cycle,
remove it from $S$,
and instead select a node from the cycle;
such a procedure will not decrease the number of predecessors
in $G - E^+(S\!)$ of any node in $S\!$).
Hence, for each node $v$ in the cycle
we can remove an outgoing edge of this node
and check wether $\Pi(G - E^+(\{v\}), p, \ell)$ is a \textit{yes}-instance
using $\texttt{alg-tree}$.
If this is the case for some node $v$ in the cycle,
we know that $(G,p,\ell)$ is a \textit{yes}-instance as well,
otherwise, it is a \emph{no}-instance.
Let us denote this procedure as \texttt{alg-connected}.

Finally, suppose that $G$ is an arbitrary functional graph.
By $c$ we denote the number of its components and
by $G_1, G_2, \dots, G_c$ the components themselves.
To solve $\Pi(G, p, \ell)$ we use a dynamic programming approach.
Let $A$ be a $c \times (\ell+1)$ array.
The values in the first row of $A$
reflect for which numbers $j \in [\ell]\cup\{0\}$
we can select $j$ nodes in $G_1$
in such a way that each has at least
$p$ predecessors after the removal of their outgoing edges,
i.e., $A[1,j] = 1$, if
$\Pi(G_1, p, j)$ is a \textit{yes}-instance,
and $A[1,j] = 0$, otherwise.
Since $G_1$ is a connected functional graph, $\Pi(G_1, p, j)$ can be decided using \texttt{alg-connected}.
Then, for each consecutive row $i \in \{2,3,\dots,c\}$,
the values reflect whether 
we can select such $j$ nodes in
$G_1,G_2,\dots,G_i$ jointly.
Thus, $A[i, j] = 1$, if there exists a $j' \in [j]$
such that $A[i-1, j'] = 1$,
and $\Pi(G_i, p, j - j')$ is a \textit{yes}-instance,
i.e., we can select $j'$ nodes in $G_1,G_2,\dots,G_{i-1}$,
and remaining $j - j'$ nodes in $G_i$
(we assume that $\Pi(G', p, 0)$ returns \textit{yes} for every graph $G'$).
Finally, $\Pi(G,p,\ell)$ is a \textit{yes}-instance, if and only if,
$A[c,\ell]=1$.\hfill \qed

\subsection*{Proof of \cref{thm:cliqueentitlement}}
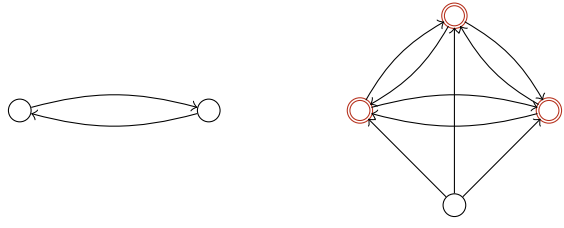
\begin{figure}[t]
	\centering
	\begin{tikzpicture}[x=5cm,y=5cm] 
		\node[e4c node] (1) at (-0.40, 0.25) {}; 
		\node[e4c node] (2) at (0.10, 0.25) {}; 
		\node[e4c node,selected] (3) at (0.50, 0.25) {}; 
		\node[e4c node,selected] (4) at (0.75, 0.50) {}; 
		\node[e4c node,selected] (5) at (1.00, 0.25) {}; 
		\node[e4c node] (6) at (0.75, 0) {}; 
		
		\path[e4c path]
		(1) edge[e4c edge, bend left=15]  (2)
		(2) edge[e4c edge, bend left=15]  (1)
		(3) edge[e4c edge, bend left=15]  (4)
		(4) edge[e4c edge, bend left=15]  (3)
		(3) edge[e4c edge, bend left=15]  (5)
		(5) edge[e4c edge, bend left=15]  (3)
		(4) edge[e4c edge, bend left=15]  (5)
		(5) edge[e4c edge, bend left=15]  (4)
		(6) edge[e4c edge]  (3)
		(6) edge[e4c edge]  (4)
		(6) edge[e4c edge]  (5)
		;
		
	\end{tikzpicture}
		\caption{An example demonstrating that \topdecay violates Clique-Entitlement (\cref{thm:cliqueentitlement}). Nodes with \textcolor{BrickRed}{red} double lines have the highest Random Walk Decay centrality.}
	\label{fig:toprank_violates}
    \vspace{-0.3cm}
\end{figure}

In \cref{fig:toprank_violates} we present a graph that demonstrates that \topdecay violates Clique-Entitlement, for $k=3$. The same instance was used in \cite{papasotiropoulos2025proportional} for showing that \toprank violates Clique-Entitlement.

In the remainder of the proof,
let us show that \absorbdecay satisfies Clique-Entitlement.
To this end, consider arbitrary graph $G = (V,E)$,
in which there exists a component $S$ that is a clique.
Let $W \subseteq V$ be an arbitrary subset of nodes of size $|W| = k < n$.
Also, let us denote the number of nodes from $S$ in $W$ by $\ell = |W \cap S|$.

We will first prove the following bounds on PageRanks of nodes in $W$
in the graph with outgoing edges of $W$ removed:
\begin{align}
	\label{eq:clique-ent:1}
	\prm^{\alpha \rightarrow 1}_{G - E^+(W)}(i) &= \nicefrac{|S|}{\ell}, \quad \mbox{for every } i \in W \cap S, \mbox{ and} \\
	\label{eq:clique-ent:2}
	\prm^{\alpha \rightarrow 1}_{G - E^+(W)}(j) &\le \nicefrac{(n - |S|)}{(k - \ell)}, \quad \mbox{for some } j \in W \setminus S.
\end{align}
Then, we will show that both formulas imply
that whenever the Clique-Entitlement condition is not satisfied, i.e.,
$\ell < \lfloor k \cdot \nicefrac{|S|}{n} \rfloor$, then
the minimum PageRank of nodes from $W$ in $G - E^+(W)$ can be increased.
Since \absorbdecay maximizes this minimum,
this will mean that \absorbdecay satisfies Clique-Entitlement.

We note that for every node $v \in V$ with a successor that does not have outgoing edges,
PageRank is also well defined for $\alpha=1$
(as the probability that the random walk returns to $v$ after visiting it
is strictly less than 1,
the expected number of visits in the walk is finite).
Thus, since we will consider only such nodes,
to simplify our calculations, we will consider PageRank with $\alpha=1$.

To prove Equation~\eqref{eq:clique-ent:1},
we will use PageRank's recursive formula,
which is an equivalent definition of PageRank~\citep{PagBriMotWin-1999-PageRank}.
The formula relates PageRank of a node
with PageRanks of the nodes from which it receives incoming edges as follows
\[
	\prm^\alpha_G(v) = 1 + \alpha \cdot \sum_{(u,v) \in E} \frac{\prm^\alpha_G(u)}{\deg^+(u)}.
\]
Consider an arbitrary unselected node in the clique, i.e., $v \in S \setminus W$.
Since in graph $G - E^+(W)$, node $v$ receives one edge from every other node in $S \setminus W$
and each such node has an out-degree of $|S|-1$,
by PageRank's recursive formula
\[
	\prm^1_{G - E^+(W)}(v) =
	1 + \frac{1}{|S|-1} \cdot \sum_{u \in S \setminus W \setminus \{v\}} \prm^1_{G - E^+(W)}(u).
\]
Now, when we sum this equation side-wise for all $v \in S \setminus W$,
on the right-hand side of the equation,
 $\prm^1_{G - E^+(W)}(u)$ for each node $u$
will appear $|S \setminus W| - 1 = |S| - \ell - 1$ times.
Thus,
\[
	\sum_{v \in S \setminus W} \prm^1_{G - E^+(W)}(v) =
	|S| - \ell + 
	\frac{|S| - \ell - 1}{|S|-1} \sum_{u \in S \setminus W} \prm^1_{G - E^+(W)}(u).
\]
Moving the sum to one side and dividing by
$1 - \frac{(|S| - \ell - 1)}{(|S|-1)} = \frac{\ell}{(|S| - 1)}$,
we get
\[
	\sum_{v \in S \setminus W} \prm^\alpha_{G - E^+(W)}(v) =
	(|S| - \ell) \frac{|S| - 1}{\ell}.
\]
Then, take an arbitrary node $i \in S \cap W$ and observe that it receives one edge from every node in $S \setminus W$.
Thus,
\[
	\prm^1_{G - E^+(W)}(i) =
	1 + \frac{1}{|S|-1} \cdot \sum_{v \in S \setminus W} \prm^\alpha_{G - E^+(W)}(v) =
	1 + \frac{|S| - \ell}{\ell} =
	\frac{|S|}{\ell}.
\]
Hence, Equation~\eqref{eq:clique-ent:1} indeed holds.

Next, let us prove Inequality~\eqref{eq:clique-ent:2}.
To this end, let us denote by $T$ the set of all predecessors of nodes in $W \setminus S$
that are not selected to $W$ themselves, i.e.,
$T = \{v \in V: \exists_{u \in W\setminus S} v \in \Pred(u)\} \setminus W$.
Then, let us sum PageRank's recursive formula sidewise,
for all nodes $v \in T$.
We obtain
\[
	\sum_{v \in T} \prm^1_{G - E^+(W)}(v) =
	|T| + 
	\sum_{(u, v) \in E : v \in T} \frac{\prm^1_{G - E^+(W)}(u)}{\deg^+(u)}.
\]
For every $u \in T$,
let $d_w(u)$ be the number of edges that go to nodes in $W$ (possibly $d_w(v)=0$),
while by $d_r(u)$ let us denote the number of its remaining outgoing edges, i.e.,
$d_r(u) + d_w(u) = \deg^+(u)$.
Observe that every node that has an outgoing edge to a node in $T$ must be in $T$ itself.
Thus, the set of edges between nodes in $T$
is exactly the same as the set of all edges counted in $d_r(u)$ for some $u \in T$.
This gives us
\[
	\sum_{v \in T} \prm^1_{G - E^+(W)}(v) =
	|T| + 
	\sum_{u \in T} \left(\frac{d_r(u)}{\deg^+(u)}\prm^1_{G - E^+(W)}(u)\right).
\]
Moving all PageRanks to the left-hand side, we get
\[
	\sum_{v \in T}  \left(\frac{d_w(v)}{\deg^+(v)}\prm^1_{G - E^+(W)}(v)\right) =
	|T|.
\]
Now, let us sum the PageRank's recursive formula for all $j \in W \setminus S$.
We get
\[
	\sum_{v \in (V \setminus S) \cap W} \prm^1_{G - E^+(W)}(v) =
	k - \ell + 
	\sum_{(u, v) \in E : v \in (V \setminus S) \cap W} \frac{\prm^1_{G - E^+(W)}(u)}{\deg^+(u)}.
\]
Observe that the set of all incoming edges to nodes in $W \setminus S$
is exactly the set of edges that are counted in $d_w(u)$ for some $u \in T$.
Thus, we get
\[
	\sum_{v \in (V \setminus S) \cap W} \prm^1_{G - E^+(W)}(v) =
	k - \ell + 
	\sum_{u \in T} \left(\frac{d_w(u)}{\deg^+(u)}\prm^1_{G - E^+(W)}(u)\right) =
	k - \ell + |T| \le
	k - \ell + (n - |S| - (k - \ell)) =
	n - |S|.
\]
Then, Inequality~\eqref{eq:clique-ent:2} follows from the pigeonhole principle.
Next, let us show that if
$\ell \le \lfloor k \cdot \nicefrac{|S|}{n} \rfloor$,
then a node in $W$ with minimum PageRank is in set $V \setminus S$.
To this end, observe that
\[
\frac{|S|}{\ell} \ge \frac{|S|}{\lfloor k \cdot \nicefrac{|S|}{n} \rfloor} \ge \frac{n}{k}.
\]
Multiplying both the nominator and the denominator by $(1 - \nicefrac{|S|}{n})$, we obtain
\[
	\frac{|S|}{\ell} \ge \frac{n(1 - \nicefrac{|S|}{n})}{k (1 - \nicefrac{|S|}{n})} = \frac{n - |S|}{k - k \cdot \nicefrac{|S|}{n}} \ge  \frac{n - |S|}{k - \lfloor k \cdot \nicefrac{|S|}{n} \rfloor} \ge \frac{n - |S|}{k - \ell}.
\]
Thus, by Equation~\eqref{eq:clique-ent:1} and Inequality~\eqref{eq:clique-ent:2},
indeed, if $\ell \le \lfloor k \cdot \nicefrac{|S|}{n} \rfloor$, then
$\min_{i \in W}\prm^1_{G - E^+(W)}(i) \le \frac{n - |S|}{k - \ell}$.
This means, that as long as the inequality is strict, i.e.,
$|S \cap W| < \lfloor k \cdot \nicefrac{|S|}{n} \rfloor$,
the minimum can be increased by removing a node with minimal PageRank from $W$
and adding to $W$ another node from the set $S$.
This means that \absorbdecay satisfies Clique-Entitlement.
\hfill \qed

\subsection*{Proof of \cref{thm:nphard}}
	First we define the decision version of the problem that we are going to prove hardness for. 
	
	\begin{table}[ht!]
		\centering
		\begin{tabular}{lp{10.8cm}}  
			\toprule
			\multicolumn{2}{c}{\textsc{Decide}-\absorbdecay } \\
			\midrule
			\textbf{Input:} & A directed graph $G=(V,E)$ and parameters $c \in \mathbb{R}_{\ge 0},k \in \mathbb{N}$.\\
			\textbf{Question:} & Does there exist a set $S \subseteq V$ such that $|S|=k$ and
			${\prm}_{G-E^+(S)}^{\alpha\rightarrow 1}(v) \geq c, \forall v\in S$?
			\\
			\bottomrule
		\end{tabular}
	\end{table}
	
	We reduce from the Independent Set (IS) problem, where,
	given an undirected graph $G' = (V', E')$ and an integer $r$,
	the goal is to determine if there exists an independent set of size $r$,
	i.e., a subset of nodes $R \subseteq V'$ such that $|R| = r$
	and no two nodes in $R$ are connected by an edge in $G'$.
	The NP-hardness holds even when $G'$ has no nodes of degree 0 or 1 \citep{dasgupta2006algorithms}.
	Given an instance of IS, $I' = (G', r)$, with $n$ nodes in $V'$ and $m$ edges in $E'$,
	we create an instance $I=(G,c,k)$ of \textsc{Decide}-\absorbdecay with graph $G=(V,E)$ as follows:
	
	\begin{itemize}
		\item We let the set of nodes $V$ contain all \textit{original} nodes from $V'$
			as well as $n+1$ \textit{new} nodes for each edge in $E'$, i.e.,
			$V = V' \cup V_{new}$, where $V_{new} = \{x_{e}^i : e \in E', i \in [n+1]\}$.
			This gives us a set $V$ of $m(n+1) + n$ nodes in total.
		\item For each undirected edge $\{u,v\} \in E'\!$,
let $E$ contain a pair of directed edges $(u,v)$ and $(v,u)$ 
			as well as an edge from each of the original nodes $u$ and $v$
			to each new node corresponding to edge $\{u,v\}$.
			Formally, we have
			$E = \bigcup_{\{u,v\} \in E'} \big(\{ (u,v),(v,u)\} \cup \{ (u, x_{(u,v)}^i), (v, x_{(u,v)}^i) : i \in [n+1]\} \big).$
		\item Finally, we set $k=r+m(n+1)$ and $c=1+\epsilon$,
			where $\epsilon$ is an arbitrary constant smaller than
			$(2 \cdot \max_{u \in V}\deg^+_G(u))^{-1}$.
	\end{itemize}
	
	The following figure depicts the used gadget for an edge $(u,v)$ of $G'$.
	The nodes with the dotted lines correspond to nodes termed as new,
	while the remaining two nodes correspond to original ones.
    
    \begin{figure}[h!]
	\begin{center}
		\begin{tikzpicture}[node distance=0.8cm, >=stealth,scale=0.58]

			\tikzset{}

			\node[e4b node] (u) at (0, 0) {$u$};
			\node[e4b node] (v) at (8, 0) {$v$};
			
			\node[e4b node,densely dotted, thick] (w1) at (4, 2) {$x_{\{u,v\}}^1$};
			\node[e4b node,densely dotted, thick] (w2) at (4, 3.5) {$x_{\{u,v\}}^2$};
			\node (dots) at (4, 4.85) {$\vdots$};
			\node[e4b node,densely dotted, thick] (wn) at (4, 6) {$x_{\{u,v\}}^{n+1}$};
			
			\path[e4c path]
			(u) edge[e4c edge] (w1)
			(u) edge[e4c edge] (w2)
			(u) edge[e4c edge] (wn)
			(v) edge[e4c edge] (w1)
			(v) edge[e4c edge] (w2)
			(v) edge[e4c edge] (wn)
			(u) edge[e4c edge, bend left=10] (v)
			(v) edge[e4c edge, bend right=-10] (u)
			;
		\end{tikzpicture}
	\end{center}
	\end{figure}
	Observe that if $\alpha \ge 1/2$,
	a node $v$ can have the value of PageRank less than $c$ in a subgraph $G''$ of $G$
	only if it does not have any incoming edges in $G''$.
	Indeed, if it has at least one incoming edge, say from $u$, then
	$\prm^\alpha_{G''}(v) \ge 1 + \nicefrac{\alpha}{\deg^+_{G''}(u)} \ge c$.
	Thus, the question of the decision variant of \textsc{Decide}-\absorbdecay\ in the instance $I$ can be equivalently expressed as follows: \textit{Does there exist a subset $S \subseteq V$ with $|S| = k$ such that $\deg^-_{G - E^+(S)}(i) > 0$ for every $i \in S$ (i.e., there is no node in $S$ that has a zero in-degree after the removal of the outgoing edges of nodes in $S\!$)?}
	We show that the positive answer to this question for $I$ is equivalent
	to the existence of an independent set of size $r$ in $I'$.
	
	First, assume that there is an independence set $R$ of size $r$ in $G'$. We will show that this implies that $S = R \cup V_{new}$, i.e.,
	the subset of the original nodes corresponding to $R$ together with all the new nodes,
	witnesses that $I$ is a yes-instance as well.
	Observe first that $|S| = r+m(n+1) = k$.
	Thus, it remains to show that every node in $S$ has a positive in-degree in the graph $G - E^+(S)$.

	Fix an arbitrary $i \in S$.
	If $i \in V_{new}$, then let $\{u,v\} \in E'$ be an associated edge.
	Since $R$ is an independent set it must be that $u \not \in R$ or $v \not \in R$.
	But this implies that in the graph $G - E^+(S)$,
	node $i$ receives an edge from $u$ or from $v$,
	so it has a positive in-degree.
	If, on the other hand, $i \in R$, then by our assumption that no node has degree of 0 in $G'$,
	we know that there is a $j \in V'$ such that $\{i,j\} \in E'$.
	Since $R$ is an independent set, it holds that $j \not \in R$.
	Thus, $i$ receives an edge from $j$ in the graph $G - E^+(S)$,
	hence it has a positive in-degree.
	This concludes the proof of the forward direction.

	For the reverse direction, assume that there exists a subset $S \subseteq V$ such that $|S| = k$
	and for every $i \in S$ it holds that $\deg^-_{G - E^+(S)}(i) > 0$.
	We will show that this implies the existence of an independent set of size $r$ in $G'$.
	Let us denote by $R$ the set of original nodes belonging in $S\!$, formally, $R := S \setminus V_{new}$.
	Since there are $m(n+1)$ new nodes in total,
	we can show that $R$ has to contain at least $r$ nodes.
	Indeed, $|R| = |S \setminus V_{new}| \ge |S| - m(n+1) = k - m(n+1) =r$.

	We will now prove that $R$ has to be an independent set in $G'$.
	For a contradiction assume that this is not the case, i.e.,
	there exists an edge $\{u,v\} \in E'$ such that $u,v \in R$.
	Observe that there must be a new node $x^i_{\{u,v\}}$, for some $i \in [n+1]$,
	that is selected to $S$.
	Otherwise, if all the new nodes associated with the edge $\{u,v\}$ were not in $S$,
	the size of $S$ would be too small, i.e.,
	$|S| \le |V| - (n+1) = m(n+1) + n - (n+1) = m(n+1) - 1 < k$.
	Since both $u$ and $v$ are in $S$,
	node $x^i_{\{u,v\}}$ has no incoming edges in $G - E^+(S)$,
 contradicting the assumption proving that $R$ is indeed an independent set.
	\hfill \qed

\end{document}